\documentclass[twocolumn, tighten]{aastex63} 

\usepackage{amsmath}
\usepackage{comment}

\newcommand{\aips}{\texttt{AIPS}}   
\newcommand{\difmap}{\texttt{Difmap}} 
\newcommand{\Rs}{R_{\rm s}} 

\newcommand{\utokyo}{Department of Astronomy, Graduate School of Science, The University of Tokyo, 7-3-1 Hongo, Bunkyo-ku, Tokyo 113-0033, Japan}
\newcommand{\naojmtk}{National Astronomical Observatory of Japan, 2-21-1 Osawa, Mitaka, Tokyo 181-8588, Japan}
\newcommand{\naojmiz}{Mizusawa VLBI Observatory, National Astronomical Observatory of Japan, 2-12 Hoshigaoka, Mizusawa, Oshu, Iwate 023-0861, Japan}

\newcommand{\haystack}{Massachusetts Institute of Technology Haystack Observatory, 99 Millstone Road, Westford, MA 01886, USA}
\newcommand{\bhi}{Black Hole Initiative, Harvard University, 20 Garden Street, Cambridge, MA 02138, USA}
\newcommand{\mpifr}{Max-Planck-Institut f\"{u}r Radioastronomie, Auf dem H\"{u}gel 69, D-53121 Bonn, Germany}

\newcommand{\iaa}{Instituto de Astrof\'{i}sica de Andaluc\'{i}a-CSIC, Glorieta de la Astronom\'{i}a s/n, E-18008 Granada, Spain}
\newcommand{\cfa}{Center for Astrophysics $|$ Harvard $\&$ Smithsonian, 60 Garden Street, Cambridge, MA 02138, USA}
\newcommand{\asiaa}{Institute of Astronomy and Astrophysics, Academia Sinica, 11F of Astronomy-Mathematics Building, AS/NTU No. 1, Sec. 4, Roosevelt Rd, Taipei 10617, Taiwan, R.O.C.}
\newcommand{\caltech}{California Institute of Technology, 1200 East California Boulevard, Pasadena, CA 91125, USA}
\newcommand{\princeton}{Princeton Center for Theoretical Science, Jadwin Hall, Princeton University, Princeton, NJ 08544, USA}
\newcommand{\bu}{Institute for Astrophysical Research, Boston University, 725 Commonwealth Avenue, Boston, MA 02215, USA}
\newcommand{\cagliari}{ Dipartimento di Fisica, Università degli Studi di Cagliari, SP Monserrato-Sestu km 0.7, I-09042 Monserrato, Italy}
\newcommand{\inaf}{INAF - Osservatorio Astronomico di Cagliari, via della Scienza 5, I-09047 Selargius (CA), Italy}
\newcommand{\niigata}{Niigata University, 8050 Ikarashi 2-no-cho, Nishi-ku, Niigata 950-2181, Japan}

\shorttitle{Jet collimation of quasar 3C\,273}
\shortauthors{Okino et al.}

\begin{document}

\title{Collimation of the relativistic jet in the quasar 3C\,273}

\correspondingauthor{Hiroki Okino}
\email{h.okino@grad.nao.ac.jp}

\author[0000-0003-3779-2016]{Hiroki Okino}
\affiliation{\utokyo}
\affiliation{\naojmtk}

\author[0000-0002-9475-4254]{Kazunori Akiyama}
\affiliation{\haystack}
\affiliation{\bhi}
\affiliation{\naojmiz}

\author[0000-0001-6988-8763]{Keiichi Asada}
\affiliation{\asiaa}

\author[0000-0003-4190-7613]{Jos\'{e} L. G\'{o}mez}
\affiliation{\iaa}

\author[0000-0001-6906-772X]{Kazuhiro Hada}
\affiliation{\naojmiz}

\author[0000-0003-4058-9000]{Mareki Honma}
\affiliation{\naojmiz}
\affiliation{\utokyo}

\author[0000-0002-4892-9586]{Thomas P. Krichbaum}
\affiliation{\mpifr}

\author[0000-0002-2709-7338]{Motoki Kino}
\affiliation{Kogakuin University of Technology $\&$ Engineering, Academic Support Center, 2665-1 Nakano, Hachioji, Tokyo 192-0015, Japan}
\affiliation{\naojmiz}

\author[0000-0003-0292-3645]{Hiroshi Nagai}
\affiliation{\naojmtk}

\author[0000-0002-7722-8412]{Uwe Bach}
\affiliation{\mpifr}

\author[0000-0002-9030-642X]{Lindy Blackburn} 
\affiliation{\cfa}
\affiliation{\bhi}

\author[0000-0003-0077-4367]{Katherine L. Bouman}
\affiliation{\caltech}

\author[0000-0003-2966-6220]{Andrew Chael} 
\affiliation{\princeton}
\affiliation{NASA Hubble Fellowship Program, Einstein Fellow}

\author[0000-0002-2079-3189]{Geoffrey B. Crew} 
\affiliation{\haystack}

\author[0000-0002-9031-0904]{Sheperd S. Doeleman} 
\affiliation{\cfa}
\affiliation{\bhi}

\author[0000-0002-7128-9345]{Vincent L. Fish}
\affiliation{\haystack}

\author[0000-0002-2542-7743]{Ciriaco Goddi}
\affiliation{\cagliari}
\affiliation{\inaf}

\author[0000-0002-5297-921X]{Sara Issaoun}
\affiliation{\cfa}
\affiliation{NASA Hubble Fellowship Program, Einstein Fellow}

\author[0000-0002-4120-3029]{Michael D. Johnson}
\affiliation{\cfa}
\affiliation{\bhi}

\author[0000-0001-6158-1708]{Svetlana Jorstad}
\affiliation{\bu}

\author[0000-0002-3723-3372]{Shoko Koyama}
\affiliation{\niigata}
\affiliation{\asiaa}

\author[0000-0003-4062-4654]{Colin J. Lonsdale}
\affiliation{\haystack}

\author[0000-0002-7692-7967]{Ru-Sen Lu}
\affiliation{Shanghai Astronomical Observatory, Chinese Academy of Sciences, 80 Nandan Road, Shanghai 200030, People's Republic of China}
\affiliation{Key Laboratory of Radio Astronomy, Chinese Academy of Sciences, Nanjing 210008, People's Republic of China}
\affiliation{\mpifr}

\author[0000-0003-3708-9611]{Ivan Mart\'{i}-Vidal}
\affiliation{Departament d'Astronomia i Astrof\'isica, Universitat de Val\`encia, C. Dr.Moliner 50, 46100 Burjassot, Spain}
\affiliation{Observatori Astron\`omic, Universitat de Val\`encia, C. Catedr\'atico Beltr\'an 2, 46980 Paterna, Spain}

\author[0000-0002-3728-8082]{Lynn D. Matthews}
\affiliation{\haystack}

\author[0000-0002-8131-6730]{Yosuke Mizuno}
\affiliation{Tsung-Dao Lee Institute and School of Physics and Astronomy, Shanghai Jiao Tong University, 800 Dongchuan Road, Shanghai, 200240, People's Republic of China}

\author[0000-0003-1364-3761]{Kotaro Moriyama}
\affiliation{Institut f\"{u}r Theoretische Physik, Goethe-Universit\"{a}t Frankfurt, Max-von-Laue-Stra$\beta$e 1, D-60438 Frankfurt am Main, Germany}
\affiliation{\naojmiz}

\author[0000-0001-6081-2420]{Masanori Nakamura}
\affiliation{National Institute of Technology, Hachinohe College, 16-1 Uwanotai, Tamonoki, Hachinohe City, Aomori 039-1192, Japan}
\affiliation{\asiaa}

\author[0000-0001-9270-8812]{Hung-Yi Pu}
\affiliation{Department of Physics, National Taiwan Normal University, No. 88, Sec. 4, Tingzhou Rd., Taipei 116, Taiwan, R.O.C.}

\author[0000-0001-9503-4892]{Eduardo Ros}
\affiliation{\mpifr}

\author[0000-0001-6214-1085]{Tuomas Savolainen}
\affiliation{Aalto University Department of Electronics and Nanoengineering, PL 15500, FI-00076 Aalto, Finland}
\affiliation{Aalto University Mets\"{a}hovi Radio Observatory, Mets\"{a}hovintie 114, FI-02540 Kylm\"{a}l\"{a}, Finland}
\affiliation{\mpifr}

\author[0000-0003-0236-0600]{Fumie Tazaki}
\affiliation{\naojmiz}

\author[0000-0003-1105-6109]{Jan Wagner}
\affiliation{\mpifr}

\author[0000-0002-8635-4242]{Maciek Wielgus}
\affiliation{\mpifr}

\author[0000-0001-7470-3321]{Anton Zensus}
\affiliation{\mpifr}

\begin{abstract}
The collimation of relativistic jets launched from the vicinity of supermassive black holes (SMBHs) at the centers of active galactic nuclei (AGN) is one of the key questions to understand the nature of AGN jets.
However, little is known about the detailed jet structure for AGN like quasars since very high angular resolutions are required to resolve these objects.
We present very long baseline interferometry (VLBI) observations of the archetypical quasar 3C\,273 at 86\,GHz, performed with the Global Millimeter VLBI Array, for the first time including the Atacama Large Millimeter/submillimeter Array.
Our observations achieve a high angular resolution down to $\sim$60\,${\rm \mu}$as, resolving the innermost part of the jet ever on scales of $\sim 10^5$~Schwarzschild radii. 
Our observations, including close-in-time High Sensitivity Array observations of 3C\,273 at 15, 22, and 43\,GHz, suggest that the inner jet collimates parabolically, while the outer jet expands conically, similar to jets from other nearby low luminosity AGN.
We discovered the jet collimation break around $10^{7}$\,Schwarzschild radii, providing the first compelling evidence for structural transition in a quasar jet.
The location of the collimation break for 3C\,273 is farther downstream the sphere of gravitational influence (SGI) from the central SMBH.
With the results for other AGN jets, our results show that the end of the collimation zone in AGN jets is governed not only by the SGI of the SMBH but also by the more diverse properties of the central nuclei.
\end{abstract}


\section{Introduction} \label{sec:intro}
Relativistic jets ejected from active galactic nuclei (AGN) are tightly collimated plasma outflows from galactic centers, known as the most energetic persistent phenomena in the Universe.
Their formation, acceleration, and collimation mechanisms constitute one of the most important questions in AGN jet physics \citep[e.g., reviewed in][]{Blandford19}.
Particularly, jet collimation processes have been studied by focusing on jet structures since they provide various information not only about the jet itself but also about their surrounding environment contributing to jet confinement \citep{Begelman94, Komissarov07, Komissarov09, Fromm18}.

The detailed nature of the jet collimation can be addressed by high-angular-resolution observations using very long baseline interferometry (VLBI) \citep[e.g.,][]{Boccardi17, Hada19}, allowing direct comparisons with special or general relativistic (SR or GR) magnetohydrodynamic (MHD) simulations \citep{Chael18a, Nakamura18, Fromm19}.
In particular, intensive studies of the M87 jet in the last decade have provided new insights about jet collimation since the M87 is the best source for investigating the global jet structures from the immediate vicinity of the central black hole (BH) to beyond the host galaxy \citep{asada_nakamura12, nakamura_asada13, Hada13, Hada16, Kim18, Walker18, Park19a, EHT_I}.
Following the aforementioned pioneering works on M87, jet shapes have been measured mainly from several nearby sources: NGC\,6251 \citep{Tseng16}, NGC\,4261 \citep{Nakahara18}, 1H\,0323+342 \citep{Hada18}, Cygnus\,A \citep{Boccardi16, Nakahara19_cygA}, NGC\,1052 \citep{Nakahara20, Baczko22}, NGC\,315 \citep{Boccardi21, Park21}, 3C\,84 \citep{Nagai14, Giovannini18}, and for a large number of jet samples collected by the long-term monitoring MOJAVE program\footnote{https://www.physics.purdue.edu/MOJAVE/index.html} \citep{Pushkarev17}.
More recently, \citet{Kovalev20} investigated jet widths from over 300 sources and found the transitions of jet shapes from ten nearby sources. They also found that the range of the transition locations is $\sim 10^{5-6}$ gravitational radii from the core, which may indicate a common property of nearby AGN jets.

The collimation properties of distant quasars are still unclear due to the difficulties in resolving the transverse jet shape. Indeed, \citet{Algaba17} systematically measured core sizes for distant radio-loud AGN to investigate the upstream jet structures. However, the core size is often unresolved even at an extremely high angular resolution with a space radio dish \citep[e.g.,][]{Gomez16}, and is often measured with high systematic uncertainties resulting from the limited angular resolution.
Although the jet structure of the flat-spectrum radio quasar (FSRQ) 4C\,38.41 was measured by \citet{Algaba19}, they could not detect significant collimation breaks because of the limited range of the measured jet scales. 
In addition, the physical parameters for this source, such as the BH mass and the jet viewing angle, have large uncertainties. Therefore, to understand the jet collimation for high-powered AGN like quasars, observations with a high angular resolution for the well-studied sources at other wavelengths are required to transversely resolve the detailed structures down to the bases of the jets.

In this paper, we study the quasar 3C\,273 (1226+023), well-known as one of the brightest extragalactic objects and the closest quasars (\citealt{Schmidt63}) with prominent jets \citep{Davis85, Conway93, Bahcall95, Jester05, Perley17}.
The 3C\,273 jets have been observed many times since their discovery because the elongated jet is a unique target for resolving down to the central (sub-) parsec (pc) scale with VLBI observations \citep{Krichbaum90, Lobanov01, Savolainen06, Kovalev16, Bruni17, Bruni21, Jorstad17, Lister19, Lister21}.
Furthermore, recent infrared interferometric observations with the GRAVITY instrument on the Very Large Telescope Interferometer (VLTI) (\citealt{GRAVITY17}) precisely estimated the BH mass and viewing angle for 3C\,273 as $M_{\rm BH}=(2.6\pm1.1)\times10^8\,M_{\odot}$ and $\theta=12^\circ\pm2^\circ$ (\citealt{GRAVITY18})\footnote{\citet{Li22} reported a new estimation of BH mass of $\sim 10^9\,M_{\odot}$ and viewing angle of $\sim 5^\circ$ for 3C\,273 during the review process of our paper. These new estimates are based on a joint analysis of spectroastrometric data in \citet{GRAVITY18} and new data from reverberation mapping with a more generalized model. We confirm that these estimates do not affect our main results and conclusions.}.
The estimated BH mass yields a linear angular relation of $1\,{\rm mas} \sim2.7\,{\rm pc} \sim1.2\times10^5\,\Rs$, which makes 3C\,273 one of the best-resolved quasars.
For these reasons, 3C\,273 is the ideal target for investigating the global structure of a quasar jet. For 3C\,273, a preliminary detection of the transition from the parabolic to conical shape has been reported in an early work of a subset of the authors in a conference proceeding \citep{Akiyama18}, based on a marginal detection of a parabolic jet from single-band data sets at 43\,GHz only covering a narrow range of the spatial area. It also lacks a careful consideration of frequency-dependent locations of radio cores at different frequencies due to the effect of synchrotron self-absorption relevant for the inner-most jet probed at millimeter wavelengths. The presence of the transition has remained inconclusive in the earlier work. 

In this paper, we report new multifrequency observations of the jet of the quasar 3C\,273 observed with several global VLBI networks. In particular, the Global Millimeter VLBI Array (GMVA) at 3.5\,mm/86\,GHz including phased Atacama Large Millimeter/submillimeter Array (ALMA) (\citealt{Matthews18}) provided the strong advantage: remarkably increasing the north-south resolution and sensitivity (see also \citealt{Issaoun19} for the first published image with GMVA+ALMA on Sgr\,A*). These observations provided us with high-fidelity imaging of the finest jet structure of 3C\,273 from the innermost sub-pc scale region with a maximum resolution of several tens of microarcseconds ($\mu$as).

The paper is organized as follows.
In Section\,\ref{sec:obs}, we describe the observations and data reduction. Images were reconstructed with the state-of-the-art regularized maximum likelihood methods as described in Section\,\ref{sec:img}, and further analyzed in Section\,\ref{sec:Image_analysis}.
We show the global jet structure of 3C\,273, including the core shift measurements and the jet collimation profile in Section\,\ref{sec:result}. The physical implications of our observations are discussed in Section\,\ref{sec:discuss}. Finally, we summarize our results in Section\,\ref{sec:summary}.
In this paper, we assume a flat $\Lambda$CDM cosmology \citep[e.g.,][]{Plank2014} with $H_{\rm 0} = 70$\,km\,s$^{-1}$\,Mpc$^{-1}$, $\Omega_{\rm M} = 0.3$ and $\Omega_{\rm \Lambda} = 0.7$, and we adopt values of $M_{\rm BH}=2.6\times10^8\,M_{\odot}$ and a viewing angle of $\theta= 9^\circ$\footnote{Previous studies have reported various values for the viewing angle of the 3C\,273 jet. \citet{Jorstad17} reported $\theta\sim6^\circ$ from VLBI monitoring at 43\,GHz. \citet{Meyer16} showed the possible range of $3.8^\circ - 7.2^\circ$. However, other studies have reported larger values of $\theta \geq \sim10^\circ$ \citep{Savolainen06, GRAVITY18}. Therefore, we assume $\theta=9^\circ$, which is broadly consistent with previously reported values of jet viewing angles.} for the 3C\,273 jet.

\begin{table*}[ttt]
 \begin{minipage}{1.00\textwidth}
 \centering
 \caption{Summary of observations of 3C\,273} \label{tab:obs_sum}
 \begin{tabular}{@{\extracolsep{\fill}}ccccccc}
    \hline \hline
    Array & Project Code & Stations & Frequency & Obs. Date   & Beam Size & Geometric Mean\\
          &              &          &   (GHz)   &(yyyy/mm/dd) & (mas$\times$mas, deg) & (mas)\\
          &              &   (1)    &    (2)    &     (3)     &          (4)          & (5)  \\ 
    \hline
    VLBA & BH151 & VLBA(10) & 1.667  & 2008/02/02 & 10.312$\times$4.460, -4.3 & 6.778 \\
    HSA  & BA122 & EB+VLBA(10) & 15.368 & 2017/03/26 & 1.025$\times$0.408, -6.3 & 0.647 \\
    HSA  & BA122 & EB+VLBA(10) & 23.768 & 2017/03/26 & 0.577$\times$0.273, -3.1 & 0.397 \\
    HSA  & BA122 & EB+VLBA(10) & 43.168 & 2017/03/26 & 0.325$\times$0.128, -6.1 & 0.204 \\
    GMVA & MA008 & AA+EB+ON+PV+YS+VLBA(8) & 86.268 & 2017/04/03 & 0.061$\times$0.052, -42.7 & 0.057 \\
    \hline\\
 \end{tabular}
 \end{minipage}
 \textbf{Notes.} (1) The stations that participated in the observations. The two-letter codes for each station are as follows; ALMA (AA), Effelsberg (EB), Onsala(ON), Pico Veleta (PV), Yebes (Ys), Brewster (BR), Fort Davis (FD), Kitt Peak (KP), Los Alamos(LA), Mauna Kea (MK), North Liberty (NL), Owens Valley (OV), Pie Town (PT), and St. Croix (SC). Ten VLBA stations (BR, FD, HN, KP, LA, MK, NL, OV, PT, and SC) are included in VLBA(10), and eight stations, excluding HN and SC, in VLBA(8). (2) The Observed central frequency. (3) The Observing date. (4) The major/minor-axis sizes and position angles of the synthesized beam with uniform weighting. (5) The beam size of an equivalent circular Gaussian with the same beam solid angle derived from the geometric mean of the major- and minor-axis beam sizes.
\end{table*}

\section{Observations and Data analysis} \label{sec:obs}

\subsection{GMVA 86 GHz} \label{subsec:GMVA}
We observed 3C\,273 with the GMVA at 86\,GHz ($\lambda\sim$ 3.5 mm) on April 3, 2017 (project code MA008), as one of the first VLBI observations with ALMA \citep[see][for a detailed description]{Goddi19}.
The phased ALMA, eight VLBA stations, and four European stations participated in our observation, as summarized in Table\,\ref{tab:obs_sum}. 3C\,273 and calibrators (3C\,279) were observed for a track of $\sim$16 hours, almost a full track for the given observing array.
ALMA participated for $\sim$5.2 hours in the middle of the track, providing baselines to both VLBA and European stations. Data were recorded at a total bandwidth of 256\,MHz per polarization, which was further subdivided into four 58\,MHz intermediate frequencies (IFs) of 116\,channels each.

Data were correlated using the DiFX correlator \citep{Dellar2011} at the Max Planck Institute for Radio Astronomy in Bonn, Germany. We note that Maunakea (MK) did not provide robust fringe detections on any of the observed sources because of the bad weather at MK throughout the GMVA campaign during Spring 2017 (see also \citealt{Issaoun19}). We also note that the visibility phases of PV baselines at each sub-band IF had an instrumental offset of 180 degrees in subsequent channels of 32\,MHz widths owing to a misconfiguration of the sub-band alignment in the correlation stage.

Initial data calibration was performed using the Astronomical Image Processing System (\aips; \citealt{Greisen03}). Preceding standard calibration, the inner-IF phase offsets in the PV baselines were corrected using a manually created bandpass (BP) table.
Visibility amplitudes were a-priori calibrated in the standard manner; visibilities at each baseline were first normalized with the available auto-correlation spectra and then scaled with the system equivalent flux densities (SEFDs) of the corresponding stations.

Phases were calibrated in a standard manner with multiple fringe fitting runs after parallactic angle correction.
First, the instrumental phases stable across the track, such as phase bandpass, inter-IF phase, and delay offsets, were corrected with a scan of the 3C\,273, providing strong detection to all stations except MK. This enabled coherent integration across the entire bandwidth and more sensitive fringe fitting by combining all IFs. Second, the delay and rate offsets of each scan were calibrated at the solution interval of the scan length, which is typically several minutes. Then, short time-scale phase rotations were fringe-fitted at solution intervals of 10\,s.

\begin{figure}[ttt]
 \centering  
 \includegraphics[width=1.0\columnwidth]{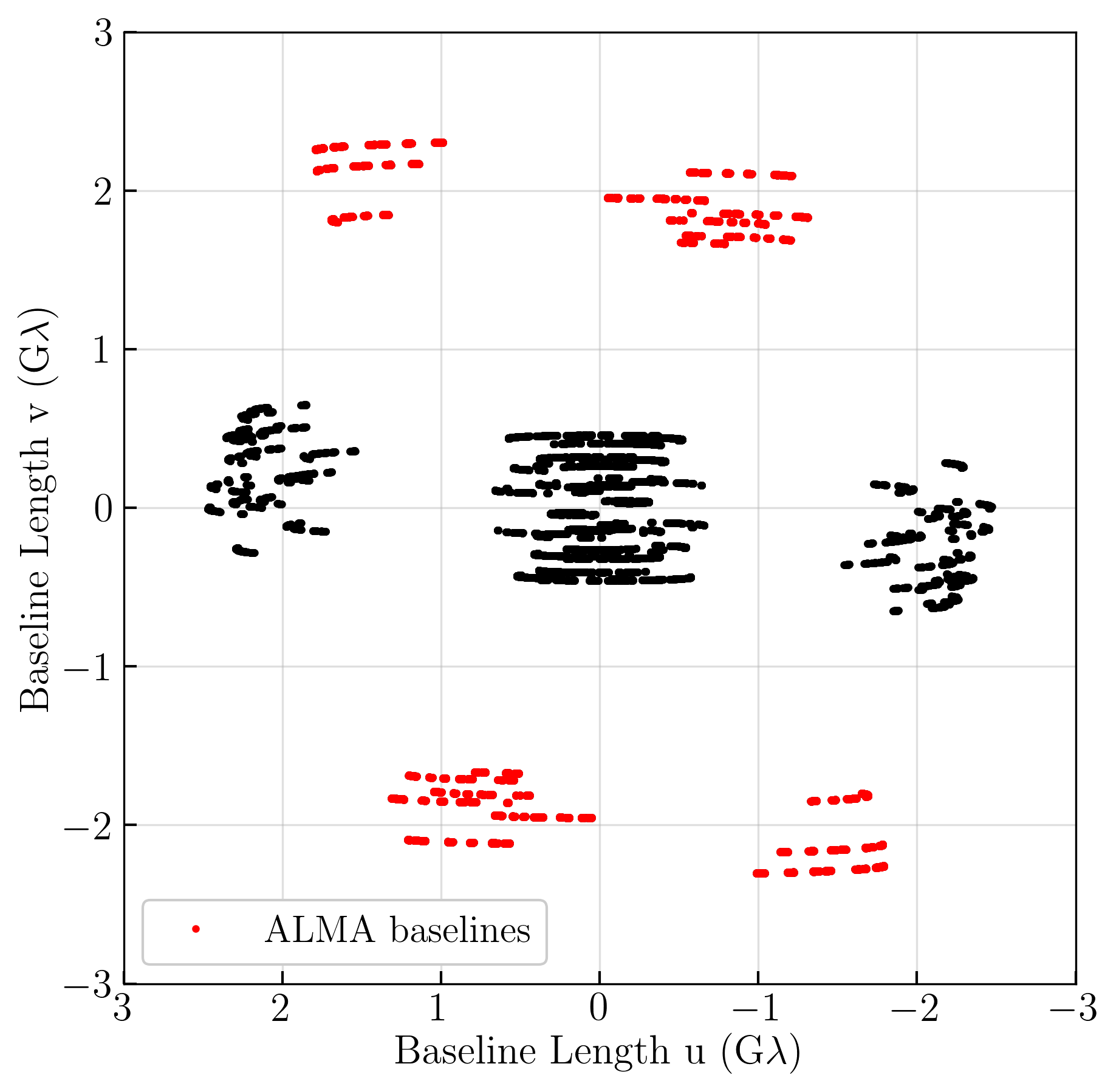}
 \caption{$uv$-coverage of the GMVA observation at 86~GHz. Baselines to ALMA are shown in red. Data are averaged at 60\,s.}
 \label{fig:uvcov86}
\end{figure}

\begin{figure}[ttt]
 \centering
 \includegraphics[width=1.0\columnwidth]{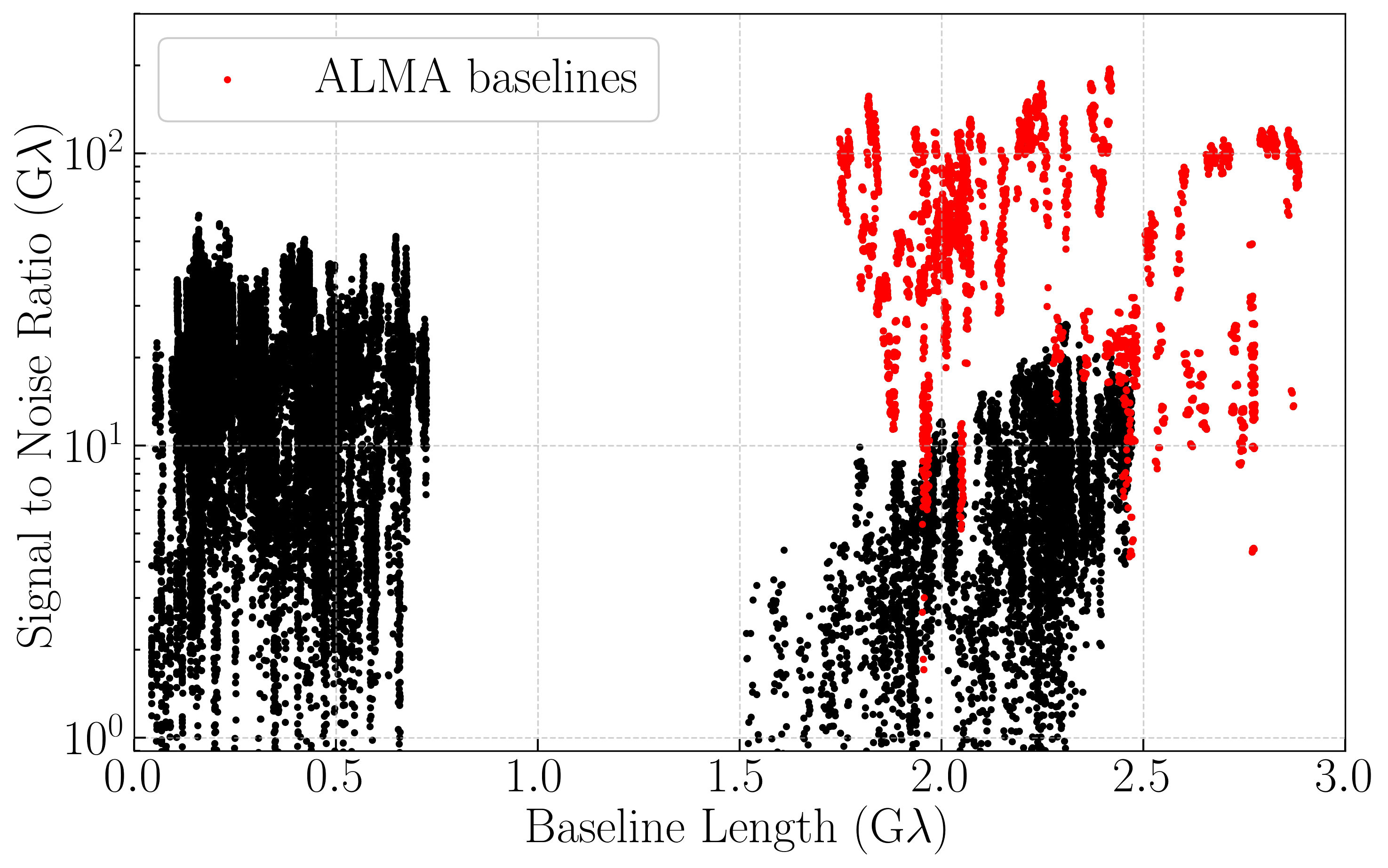}
 \caption{The SNRs of visibilities on 3C\,273 as a function of projected baseline length. Data are averaged at 60\,s. Baselines to ALMA are shown in red.}
 \label{fig:snr86}
\end{figure}

The new addition of ALMA to GMVA has significantly improved the overall array performance.
Figure\,\ref{fig:uvcov86} shows the $uv$-coverage for our GMVA observation of 3C\,273.
The long baselines beyond $\sim$1.5\,G$\lambda$ in north-south (N-S) direction correspond to ALMA.
The phased ALMA has improved the angular resolution of the GMVA observations in the N-S direction by more than a factor of two (see Section\,\ref{sec:GMVA_image}).
Furthermore, the ALMA provides very high SNR detection on long baselines to both VLBA and European stations at $\sim1.5- 2.0\,{\rm G}\lambda$ (Figure\,\ref{fig:snr86}).
The sensitive ALMA secured the detection of fringes to most of the stations in the array while it was participating the observation.

\subsection{HSA 15/22/43 GHz} \label{subsec:HSA}
We performed multifrequency observations of 3C\,273 at 15, 22, and 43\,GHz with the High Sensitivity Array (HSA) on March 26, 2017, eight days before our GMVA observations (project code: BA122).
The observing array consisted of the ten VLBA stations and the 100-m Effelsberg (EB) telescope in Germany, as summarized in Table\,\ref{tab:obs_sum}.
Full-track observations were performed for $\sim$11\,h, of which EB participated in the first $\sim$3.4\,h. Data were recorded at dual circular polarizations with four 64\,MHz IFs subdivided into 256\,channels, providing an aggregate of 256\,MHz per polarization. Data were correlated using the VLBA DiFX correlator.

The initial data calibration for the HSA data was performed using \aips.
We note that the EB baselines in several IFs showed lower visibility amplitudes and larger scattering in visibility phases due to the bandpass issue, which cannot be calibrated in this stage. These data were removed prior to the data calibration. Data were calibrated in a standard manner similar to the GMVA data sets described in Section\,\ref{subsec:GMVA}; amplitudes were a-priori calibrated, and then phases were calibrated with fringe fitting. Fringes were detected on all stations, and similar baseline coverages were obtained at all three observing frequencies.

\subsection{VLBA 1.7 GHz} \label{subsec:VLBA}
To complement our data on the larger-scale jet structure, we reduced an archival VLBA 1.7\,GHz data set of 3C~273 observed in February 2008 (project code: BH151). The data set has one of the best hour-angle coverage among existing archival data sets at 1.7\,GHz\footnote{We cross-checked archival data of a more close-in-time VLBA observation at 1.6\,GHz in 2014 with the experiment code of BG216H. We obtained a jet width profile consistent with the data sets used in this paper, confirming that the jet width of the 1.6\,GHz jet does not change significantly over the time scale of 10 years.}. Data were reduced in a standard manner in \aips. Fringes were detected on all ten VLBA stations, as summarized in Table\,\ref{tab:obs_sum}.

\section{Imaging} \label{sec:img}

\begin{table*}[ttt]
 \begin{minipage}{1.00\textwidth}
 \centering
 \caption{List of Imaging Parameters}
 \label{tab:obs_imaging}
 \begin{tabular}{cccccccc}
    \hline \hline
    Freq.     & Sys err & $\ell_{1}^{w}$ & TV & TSV & MEM & No. of Total & No. of Top-set\\
    (GHz)     & (\%)    &                &    &     &     & Parameters   & Parameters  \\
    \hline
    1.6& 0, 1 &  $1$ & 1 & $0, 10^{-2}, 10^{-1}, 1, 10$ & $0, 10^{-2}, 10^{-1}, 1$ & 40 & 15 \\
    15 & 0, 1 &  $10^{-1}, 1, 10$ & 1 & $0, 10^{-2}, 10^{-1}, 1, 10$ & $0, 10^{-2}, 10^{-1}, 1, 10$ & 150 & 45\\
    22 & 0, 1 &  $10^{-1}, 1, 10$ & 1 & $0, 10^{-2}, 10^{-1}, 1, 10$ & $0, 10^{-2}, 10^{-1}, 1, 10$ & 150 & 30\\
    43 & 0, 1 &  $10^{-1}, 1, 10$ & 1 & $0, 10^{-2}, 10^{-1}, 1, 10$ & $0, 10^{-2}, 10^{-1}, 1, 10$ & 150 & 35\\
    86 & 0, 1 &  $10^{-1}, 1$ & 1 & $0, 10^{-2}, 10^{-1}, 1$ & $0, 10^{-2}, 10^{-1}, 1$ & 64 & 24\\\hline
 \end{tabular}
 \end{minipage}
\end{table*}

We reconstructed multifrequency images of 3C\,273 from the GMVA, HSA, and VLBA data sets presented in Section\,\ref{sec:obs} with {\it regularized maximum likelihood} (RML) methods implemented in {\tt SMILI} \citep{Akiyama17a, Akiyama17b}.
Our GMVA and HSA arrays are heterogeneous and were expected to have more residual calibration errors at higher frequencies, such as 43 and 86\,GHz. Furthermore, our GMVA observations had sparse $uv$-coverages lacking intermediate baselines as shown in Figure\,\ref{fig:uvcov86}. The traditional iterative hybrid mapping using CLEAN \citep[e.g.,][]{Hogbom1974} and self-calibration implemented in popular packages such as \aips ~and \difmap ~ (\citealt{Shepherd97}), is generally more challenging than for the lower frequency data sets and does not offer the flexibility. RML methods provide a more flexible imaging framework, by directly using various types of data sets including robust closure quantities \citep[e.g.,][]{Thompson17} and a wider range of assumptions for the source images.

We employed an imaging approach inspired by recent EHT imaging of M\,87 \citep{EHT_IV} that explored a wide range of parameters, implying assumptions of source images, and assessed uncertainties in images and their deliverables more conservatively. Using a scripted RML imaging pipeline, we explored the distribution of images giving satisfactory fit to the data, which were all further used in the image analysis in Section \ref{sec:Image_analysis}. We briefly introduce the RML method in Section\,\ref{sec:imaging_Background}. Then, we describe the imaging pipeline used in our imaging of 3C\,273 in Section\,\ref{sec:imaging_pipeline}, and the details of the imaging parameter survey in Section\,\ref{sec:imaging_prm_survey}.

\subsection{RML methods} \label{sec:imaging_Background}
RML methods are a new class of imaging techniques that were conceived to overcome many of the technical challenges of millimeter VLBI imaging, particularly with the EHT.
RML imaging takes a forward-modeling approach inspired by Bayesian statistics, directly solving for an image without using a dirty beam or dirty map. RML methods have been demonstrated to improve the overall quality of image reconstruction not only for synthetic observations \citep[e.g.,][]{Honma14,Chael16, Chael18a, Akiyama17a, Akiyama17b, Kuramochi18} but also for actual interferometric measurements with VLBI arrays \citep[e.g.,][]{Issaoun19, EHT_IV, Kim20, Janssen21} and connected interferometers \citep[e.g.,][]{Matthews18, Yamaguchi20}. 

RML methods derive a reasonable or conservative image from an infinite number of images consistent with given interferometric measurements by solving for an image that minimizes the sum of data consistency metrics, such as $\chi^2$ terms, and regularization functions mathematically describing prior assumptions for the source morphology.
This framework allows a flexible choice of input data, for instance, the direct use of closure quantities free from antenna-based calibration errors \citep[e.g.,][]{Chael16,Chael18a,Akiyama17a, Blackburn20}, and further inclusion of various observing effects such as systematic non-closing errors \citep[][]{EHT_IV}.
Popular regularization functions include $\ell_1$-norm, total variation (TV) and total squared variation (TSV), enforcing sparsity in some basis of the image \citep{Honma14,Ikeda16,Akiyama17a,Akiyama17b,Kuramochi18} and the information entropy of the image \citep[e.g.,][]{Chael16}.
By combining various regularization functions, RML methods can explore a wide range of images consistent with the data, often leading to reconstruction with more reasonable assumptions of the target source \citep[e.g.][]{Akiyama17a, Akiyama17b, EHT_IV}. Further technical and mathematical details of the RML approaches adopted in this study are described in \citealt[][]{EHT_IV}.

\subsection{Imaging Pipeline} \label{sec:imaging_pipeline}
We used a scripted pipeline in Python for our RML imaging with {\tt SMILI}. In the pipeline, 3C\,273 images were reconstructed from AIPS-calibrated data (see Section\,\ref{sec:obs}) by utilizing weighted-$\ell _1$, TV, TSV and MEM regularizers \citep[see Appendix A of][for mathematical definitions]{EHT_IV}.
The pixel size of the image is typically set to one-tenth of the mean full width at half maximum (FWHM) size of the uniform-weighted synthesized beam summarized in Table\,\ref{tab:obs_sum}.
Prior to imaging, the $uv$-coordinates of the data were rotated by $45^{\circ}$ counterclockwise, resulting in the same rotation of the image-domain axis, and the horizontal axis approximately aligned with the jet direction.
This allows to use a narrower rectangular image field of view and minimizes the computational cost of imaging. Then, the visibilities were coherently averaged for 60\,sec, after manual flagging outliers. If specified, systematic errors were added in quadrature to the thermal noise of the time-averaged complex visibilities to account for non-closing errors. We explored 0 or 1\% of non-closing errors within the range expected for polarization leakages \citep[e.g.,][]{Zhao22}.

Our imaging procedure is iterative, with four stages of imaging and self-calibration. In the first stage, the initial image is set to a circular Gaussian with the size corresponding to the geometric mean of the major and minor-axis sizes of the synthesized beam at each observed frequency (see Table\,\ref{tab:obs_sum}). The subsequent stages of imaging started with the final image of the previous stage convolved with the above circular Gaussian. The first two stages begin with the initial image and use visibility amplitudes, log closure amplitudes, and closure phases for imaging. At each stage, the visibility amplitudes of all baselines added a 5\% error in quadrature to the thermal noise to account for the amplitude calibration uncertainties. In the final two stages, the imaging uses complex visibilities and closure quantities. 

\subsection{Imaging Parameter Survey} \label{sec:imaging_prm_survey}
Using the pipeline described in Section\,\ref{sec:imaging_pipeline}, we explored a wide range of parameters, as summarized in Table \ref{tab:obs_imaging}.
At each frequency, the parameter survey examined several tens to a few hundreds, of parameter sets with four to five parameters: potential systematic non-closing errors (denoted as Sys err) and the regularization parameters for the $\ell_{1}^{w}$, TV, TSV, and MEM regularizers.
Among the sets of the imaging parameters explored, we selected ``top sets'' of the parameters that provide reconstructed images with reasonable fits to data. The distributions of the corresponding top-set images allow us to assess and identify the morphologies commonly seen and insensitive to the imaging choices as well as their uncertainties among the reconstructed images consistent with our interferometric measurements.

We started imaging with the HSA data at 15\,GHz. We adopted the published VLBA image at 15\,GHz and Stokes $I$ on May 25, 2017 (project code BL229AH) from the MOJAVE program \citep{Lister18}, which was the closest epoch to our HSA observations, as the softmask of the imaging region with $\ell_{1}^{w}$ regularization. This allows noise suppression outside of the area, which has historically had no significant emission \footnote{The typical jet proper motion of 3C\,273 is a few mas/yr, which does not change the overall jet emission structure during a few months}. Then, the best-fit image among the top-set 15\,GHz images is adopted as the soft mask for subsequent 22\,GHz image processing to maximize the consistency between adjacent frequencies. We processed data at higher frequencies (43 and 86\, GHz) in the same manner using best-fit images at the adjacent lower frequencies.
For L-band data, we used the L-band VLBA image in \citet{Akiyama18} as its soft mask.

We selected the top-set images with good fits to the data using $\chi^{2}$ statistics. For the selection of top-set parameters, we adopted the minimum threshold of 1.5 for the reduced $\chi^{2}$ values of full complex visibilities, amplitudes, and closure quantities of self-calibrated images. Consequently, fifteen to forty-five images were selected at each frequency, as summarized in Table\,\ref{tab:obs_imaging}, and used in post-imaging analysis.

\subsection{Total flux scaling for GMVA 86 GHz} \label{sec:flux_scale}
The overall scaling of visibility amplitudes strongly depends on the accuracy of the {\it a-priori} calibration of visibility amplitudes based on measurements of the system-equivalent flux density at each station, which often have large systematic errors in high-frequency VLBI observations above $\sim$86\,GHz \citep[see e.g.,][for previous GMVA observations]{Koyama16, Kim18}.
We assumed Effelsberg (EB) and Pico\,Veleta (PV), often considered as reliable stations for a-priori calibrations, as the reference stations for overall gain scaling \citep[see e.g.,][]{Angelakis15, Fuhrmann16, Agudo18b}. We scaled the total flux density of our GMVA 86\,GHz images such that the median gain amplitude of EB and PV is unity. 
The derived scaling factor is $\sim 1.46$, and we obtained the scaled total flux (median in the top-set) is $\sim3.1$\,Jy.
The scaling of amplitudes and the total flux density described here will not affect our main results based on the collimation profile measured from the source morphology (Section\,\ref{sec:jet_measure}).
We note that our 86\,GHz images show reasonable spectral indices to images at lower frequencies (see Section \ref{sec:spec_index}).

\section{Image analysis} \label{sec:Image_analysis}
With all top-set images from the imaging parameter survey (Section\,\ref{sec:imaging_prm_survey}), we investigate the jet profile as a function of distance from the central black hole. First, we identified the position of the central black hole using the core shift effect as described in Section \ref{sec:core_measure}. Then, the jet radii at each frequency were measured, as described in Section \ref{sec:jet_measure}.

\subsection{Core Shift Measurements} \label{sec:core_measure}
Core shift is a positional shift of the radio core between two different observing frequencies due to the frequency dependence of the optical depth of the synchrotron self-absorption \citep{Blandford79}.
In our study, we employed the widely-used self-referencing method \citep[e.g.][]{Lobanov98, O'Sullivan09, Pushkarev12, Fromm13, Hada18} to measure the core shift of 3C\,273, based on a well-vetted assumption that the emitting regions from the extended jet are optically thin features, and their positions do not change at different frequencies. From this assumption, one can obtain the core offset values between two images at different frequencies after aligning the corresponding optically thin emission regions. We only used the images at 15, 22, 43, and 86\,GHz observed within a week (see Table\,\ref{tab:obs_sum}). The offsets were measured for four combinations (15/22, 22/43, 43/86, 15/43\,GHz)\footnote{The other two (i.e., 15/86 and 22/86 GHz) pairs were omitted due to the lack of extended emission commonly seen in both frequencies.}. For each frequency pair, we derived the offsets using all combinations of the top-set images and adopted their mean values as the measured offsets.

First, we needed to identify the core position for each jet image to compare them at different frequencies. We decided on the locations of some components in the upstream regions by several circular Gaussian model fittings using \difmap. We applied this analysis to all the top-set images and identified four to seven components for the 15-86\,GHz images. We set the core at the most upstream component for each image.

To obtain the core offset between two frequencies, we used a two-dimensional cross-correlation of the optically thin emission regions in the jet images following the widely used method described in \citet{Croke08}.
First, the pixel size of both images was set to a twentieth of the restoring beam size at the lower frequency, which was much smaller than the angular resolution at both frequencies. Next, we convolved both images with the restoring beam at the lower frequency. After masking the optically thick core regions of both images, we computed the cross-correlation coefficient $r_{xy}$ defined by
\begin{equation} \label{eq_crosscorr}
  r_{xy} \\
  = \frac{\Sigma^n_{i=1}\Sigma^n_{j=1}(I_{\nu1,ij}-\bar{I}_{\nu1})(I_{\nu2,ij}-\bar{I}_{\nu2})}{\sqrt{\Sigma^n_{i=1}\Sigma^n_{j=1}(I_{\nu1,ij}-\bar{I}_{\nu1})^2\Sigma^n_{i=1}\Sigma^n_{j=1}(I_{\nu2,ij}-\bar{I}_{\nu2})^2}}
\end{equation}
where $I_{\nu,ij}$ is the intensity at pixel $(i,j)$ at frequency $\nu$, and $\bar{I}$ is the averaged intensity in the calculated region. The cross-correlation coefficient $r_{xy}$ is a function of the relative positional offset $(\Delta x, \Delta y)$. The positional shift between two images at frequencies $\nu_1$ and $\nu_2$ is given by the location of the global maximum $r_{xy}$. The core offset was defined as the remaining positional offset of the peaks after aligning the corresponding optically thin emission regions.

\subsection{Jet Profile Measurements} \label{sec:jet_measure}
We investigated the jet collimation in 3C\,273 using multifrequency images.
Prior to analysis, we cut off the brightness distribution for all the images below the rms values (see Section\,\ref{sec:GMVA_image}) at each frequency to remove the noise outside the jet regions.
Then, the jet radius was measured as a function of distance from the core at each frequency in two steps.

First, we measured the position angle (PA) of the jet at each distance from the core. Following \citet{Pushkarev17}, we took a circular slice of the image centered at the core with a radius corresponding to each distance and adopted the intensity-weighted centroid as the PA of the jet at the corresponding distance.

After deriving the PA, we measured the jet radius at each distance. Because the jet bends on milliarcsecond scales, we examined two ways of slicing the image to measure the jet radius. The first one is taking a slice perpendicular to the PA at each distance, and the other is slicing the image perpendicular to the local tangent line of the PA profile, following \citet{Pushkarev17}. The effective jet radius was measured by taking the second moment $\sigma_{\rm w}$ of the cross-section along with each slice at each distance. Then, the FWHM width of the jet was obtained by ${\rm FWHM}=2\sqrt{2\ln{2}}\sigma_{\rm w}$. We derived the deconvolved FWHM width ($w$) defined as $w=\sqrt{{\rm FWHM}^2 - \theta_{\rm beam}^2}$, correcting the blurring effects with the restoring beam at the FWHM of $\theta_{\rm beam}$\footnote{Although interferometric imaging with RML, in principle, does not require a post-imaging beam convolution to obtain piecewise smooth images, the raw image may contain some over-resolved features that are not strongly supported by data. To reduce potential bias from such features, in this work, we adopt the conventional approach that derives de-convolved widths from ones measured from beam-convolved images widely used with CLEAN imaging.} (see Table\,\ref{tab:obs_sum}). Finally, the jet radius $r$ at each distance was obtained from half size of the deconvolved FWHM width (i.e., $r=w/2$). 

The jet radius profile was measured with an interval of half of the restoring beam size (see Table \ref{tab:obs_sum}). We only accept the jet radius measurements satisfying the following three criteria: (i) the FWHM is larger than $\theta_{\rm beam}$ so that the vertical jet emission is well resolved, (ii) the distance from the radio core is greater than $2\times\theta_{\rm beam}$ to conservatively remove the effect of the unresolved core, and (iii) the peak intensity of the slice is three times larger than the mean of the residual rms noise derived from all top-set images (see Section\,\ref{sec:result}) at each frequency except for 22\,GHz. At 22\,GHz, instead, the threshold was set to be nine times larger than the mean residual rms noise to make reliable measurements only at bright locations in the sparse intensity distribution (see the image in Figure\,\ref{fig:HSA&VLBA_mean_image}). The jet radius profile at each frequency was measured for all top-set images with the above two slicing methods. Then, the mean and standard deviation of the jet radius at each distance and frequency were adopted as the corresponding jet radius and its uncertainty, respectively.

\section{Results} \label{sec:result}

\subsection{GMVA image} \label{sec:GMVA_image}

\begin{figure*}[ttt]
 \centering
 \includegraphics[width=0.75\textwidth]{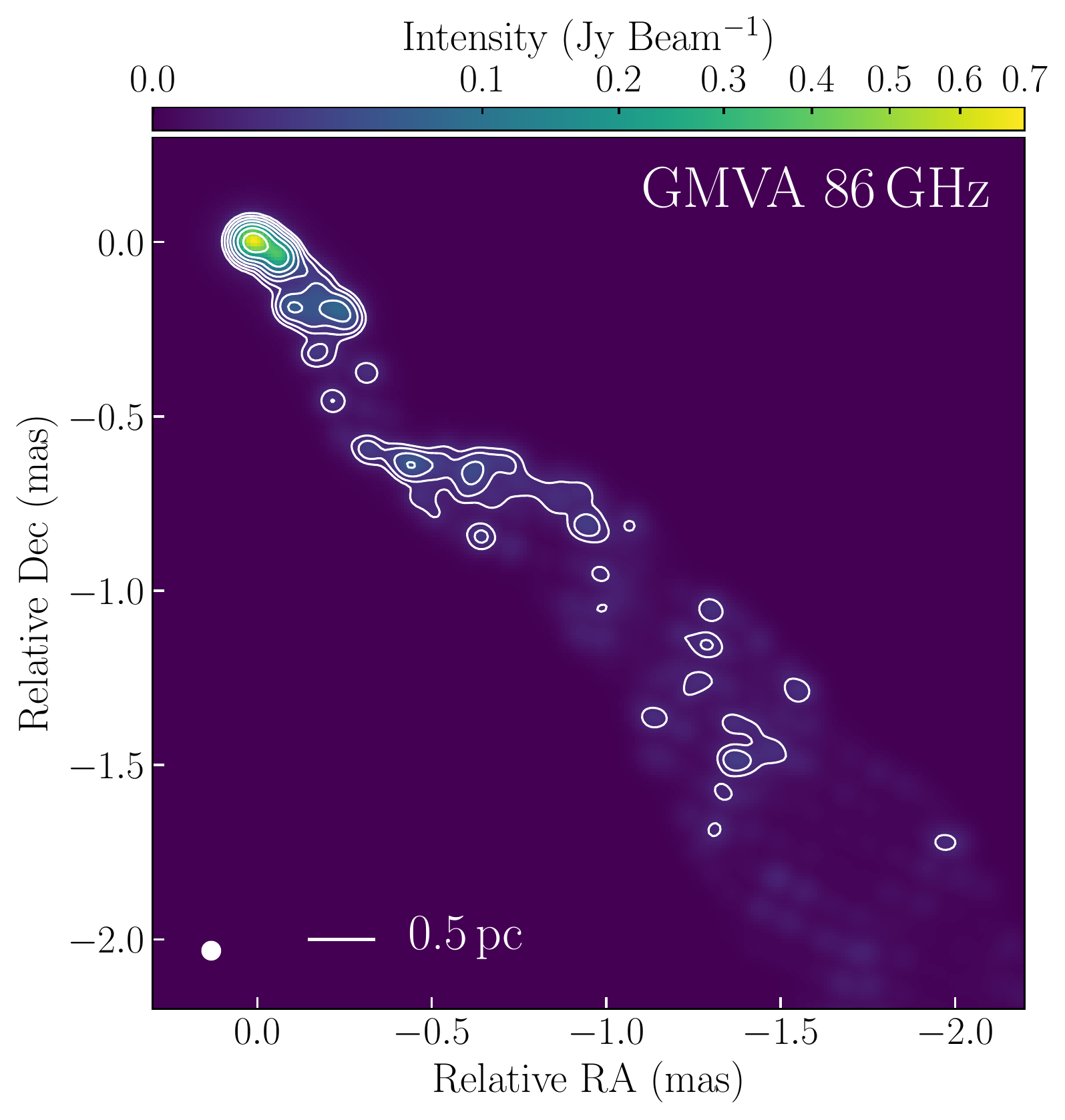}
 \caption{The inner-jet images of 3C\,273 obtained with the GMVA observation including ALMA at 86\,GHz. Here we show the mean of the top-set reconstructions restored at the resolutions in uniform). The image restored with a circular beam of 57\,$\rm \mu$as (labeled in the bottom left corner) corresponding to the geometric mean of the uniform-weighting beam of $61 \times 52\,\rm \mu$as. The peak intensity is 0.69\,Jy/beam. The contours are multiplies by the factor of 2, from the lowest contour level of $6.5$\,mJy/beam.}
 \label{fig:gmva_mean_image}
\end{figure*}

In Figure\,\ref{fig:gmva_mean_image}, we show the mean total intensity image of the 3C\,273 jet with GMVA at 86\,GHz derived by averaging all top-set reconstructions (24 images) obtained from the imaging parameter survey in Section\,\ref{sec:imaging_prm_survey}. 
Here we show the mean image restored at two characteristic resolutions of our GMVA observations equivalent to uniform weighting\footnote{The restoring beam is arbitrary for RML methods, given the nature of methods that the synthesized beam is not used for deconvolution. Data weighting is equivalent to the natural weighting of CLEAN imaging since data are only weighted by thermal noise and not by $uv$-density, while the reconstructed images are known to have a high fidelity at the resolution comparable to, or higher than, uniform weighting.}.
Our new image provides the sharpest view of the inner jet at 86\,GHz thanks to the critical addition of ALMA. Without ALMA, the corresponding beam sizes will be $265\times50\,\rm{\mu as}$ in uniform weighting providing significantly worse resolutions along the NS direction by a factor greater than 2.5. The spatial resolution of our observations is also higher than previous observations, such as the high-sensitivity VLBA + GBT observations in \citet{Hada16} with a resolution of $340\times110\,\rm{\mu as}$ at PA=$-10^{\circ}$. 
The high-resolution of our new observation allows us to resolve the vertical structure of the inner jet within a few mas from the core at 86\,GHz for the first time.

\begin{figure}[ttt]
 \centering
 \includegraphics[width=1.0\columnwidth]{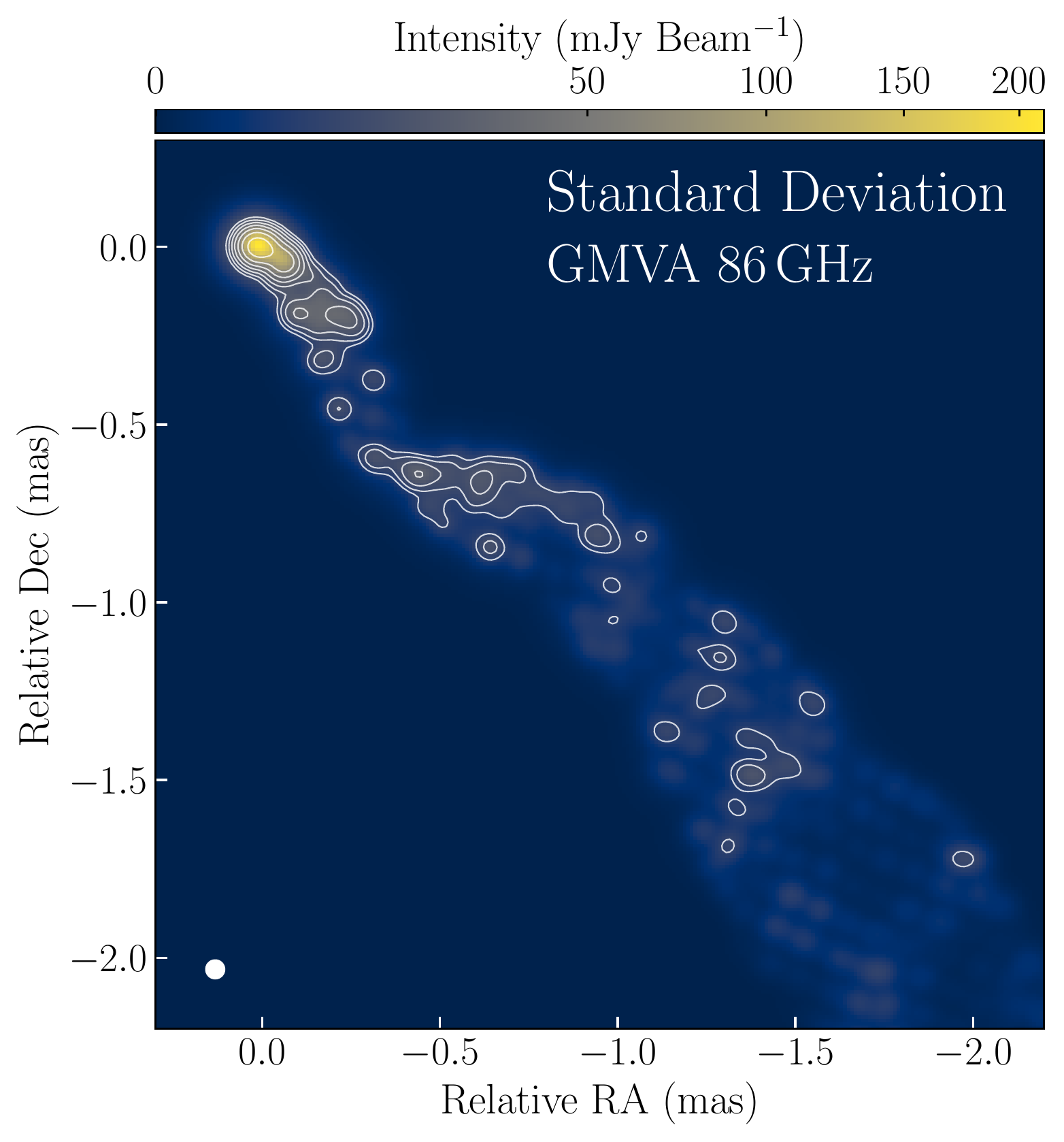}
 \caption{Standard deviation of the top-set reconstructions of GMVA 86\,GHz restored with the circular Gaussian beam of $57\,\rm{\mu as}$ shown by the white circle in the bottom left corner. The contours represent the mean image shown in the left panel of Figure\,\ref{fig:gmva_mean_image}.}
  \label{fig:gmva_uncert_image}
\end{figure}

To assess the image uncertainty quantitatively, we obtained residual images using the top-set reconstructions (not blurred) as the image models in the \difmap~software. We calculated the image rms values for each residual in the top-set, and derived the mean value of $\sim3$\,mJy/beam as the representative image uncertainty at 86\,GHz with natural weighting scheme. We then determined the lowest contour level for the uniform weighing beam to be the same dynamic range ($\sim100$) as the one for natural weighting.

In addition, the final top-set reconstructions have some variations in the image domain depending on the imaging parameters. Figure\,\ref{fig:gmva_uncert_image} shows the standard deviation image of GMVA 86\,GHz across the top-set reconstructions. The brightest core shows the largest ($\sim200\,\rm{mJy/beam}$), while the faint and extended emissions show smaller variation. These results suggest that the variability in the image domain, especially in the bright region, is dominated by the choice of imaging parameter rather than the rms of the residual images.

The median value of the scaled total flux densities among the top-sets of 86\,GHz images is $\sim3.1$\,Jy (Section\,\ref{sec:flux_scale}), corresponding to $\sim 30\%$ of the arcsecond-scale core flux density measured with the ALMA in Band\,3 during this GMVA campaign \citep[][]{Goddi19}.
The extended jet emission at 3mm is broadly consistent with the 7mm jet implying optically thin spectra (see Section\,\ref{sec:spec_index}), suggesting that the rest of the ALMA core flux density has been resolved out at the shortest VLBI baselines with fringe spacings equivalent to a few milliarcseconds. 

\subsection{HSA and VLBA images} \label{sec:HSA_VLBA_image}

\begin{figure*}[ttt]
 \centering
 \includegraphics[width=1.02\textwidth]{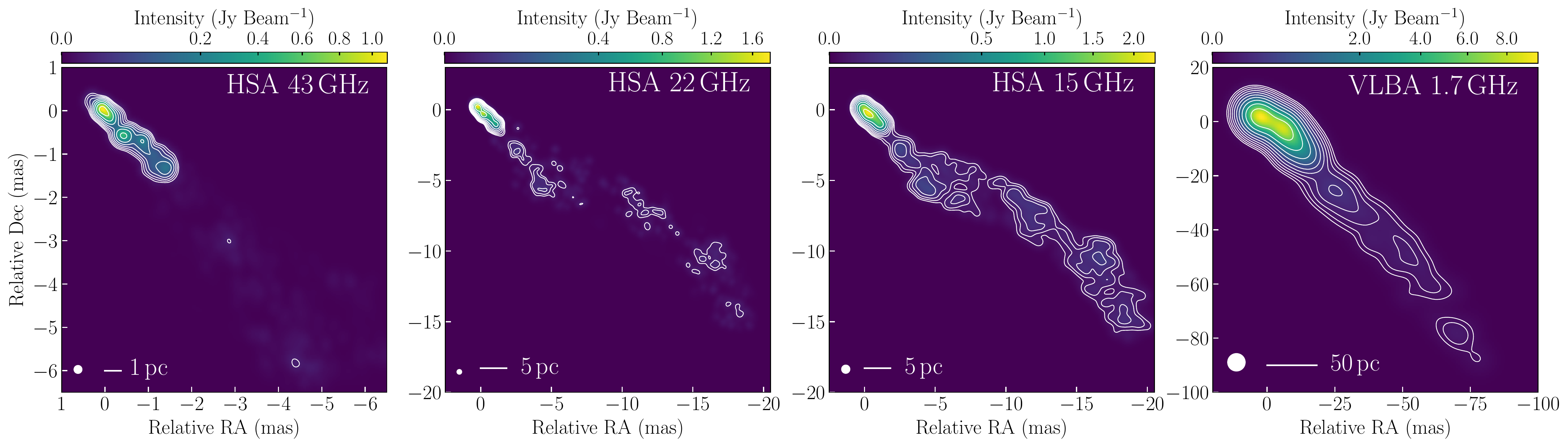}
 \caption{Multifrequency images of 3C\,273 jets with the HSA at 43, 22, and 15\,GHz and the VLBA at 1.7\,GHz. Each panel shows the mean top-set image at each frequency restored by the circular Gaussian beam with the beam-solid angle of the uniform-weighting beam size in Table\,\ref{tab:obs_sum}.
 The lowest contour levels are $6.5$\,mJy/beam at 43\,GHz, $8.2$\,mJy/beam at 22\,GHz, $7.1$\,mJy/beam at 15\,GHz, and $11.6$\,mJy/beam at 1.7\,GHz.The contours for all images are multiplied by a factor of 2. The lowest contour level of each image is estimated from the mean residual image rms of the top-set reconstructions.}
 \label{fig:HSA&VLBA_mean_image}
\end{figure*}

In Figure\,\ref{fig:HSA&VLBA_mean_image}, we show lower-frequency images of 3C\,273 observed with the HSA and VLBA, which were all averaged over the top-set reconstructions similar to the GMVA images in Figure\,\ref{fig:gmva_mean_image}. 
The overall jet morphology is consistent with previous ground-based (MOJAVE: \citealt{Lister18}; VLBA-BU-Blazar\footnote{\url{http://www.bu.edu/blazars/VLBAproject.html}}: \citealt{Jorstad17}) and Space VLBI observations ({\it Radio Astron}: \citealt{Bruni17, Bruni21}). Regardless of the observing frequency, the 3C\,273 jet extends in the southwest direction with multiple bright knots whose relative locations are consistent between different frequencies.

The HSA 43\,GHz image shows the inner jet structure within $\sim2$\,mas, and extended two components around 4 and 7\,mas from the core. These overall morphology is consistent with a close-in-time image in the database of the VLBA-BU-Blazar program observed a week ago (March 19, 2017). 

The HSA 15 and 22\,GHz images show the same regions with a bright core and extended jet emissions up to $\sim25$\,mas from the core.  
One can see that the direction of the jet changes at $\sim10$\,mas from the core, with the PA changing from $\sim-140^{\circ}$ (upstream) to $\sim-120^{\circ}$ (downstream). 
In the jet, several components are located at the same position at both frequencies. However, the faint jet component is not detected at 22\,GHz unlike at 15\,GHz, due to the optically thin synchrotron emission from the jet (see Section\,\ref{sec:spec_index}) and also the limited $uv$-coverage for short baselines at 22\,GHz compared with 15\,GHz. 

In our VLBA 1.7\,GHz image, the jet emissions downstream of $\sim20\,\rm{mas}$ are reconstructed with our imaging. The image is similar to the 3C\,273 jet shown in \citet{Kovalev16} at the same frequency band. One can see the bright component at $\sim40\,\rm{mas}$ and the extended jet emission at $\sim100\,\rm{mas}$ from the core.

\subsection{Spectral Index Maps}
\label{sec:spec_index}

\begin{figure*}[ttt]
 \centering
 \includegraphics[width=1.0\textwidth]{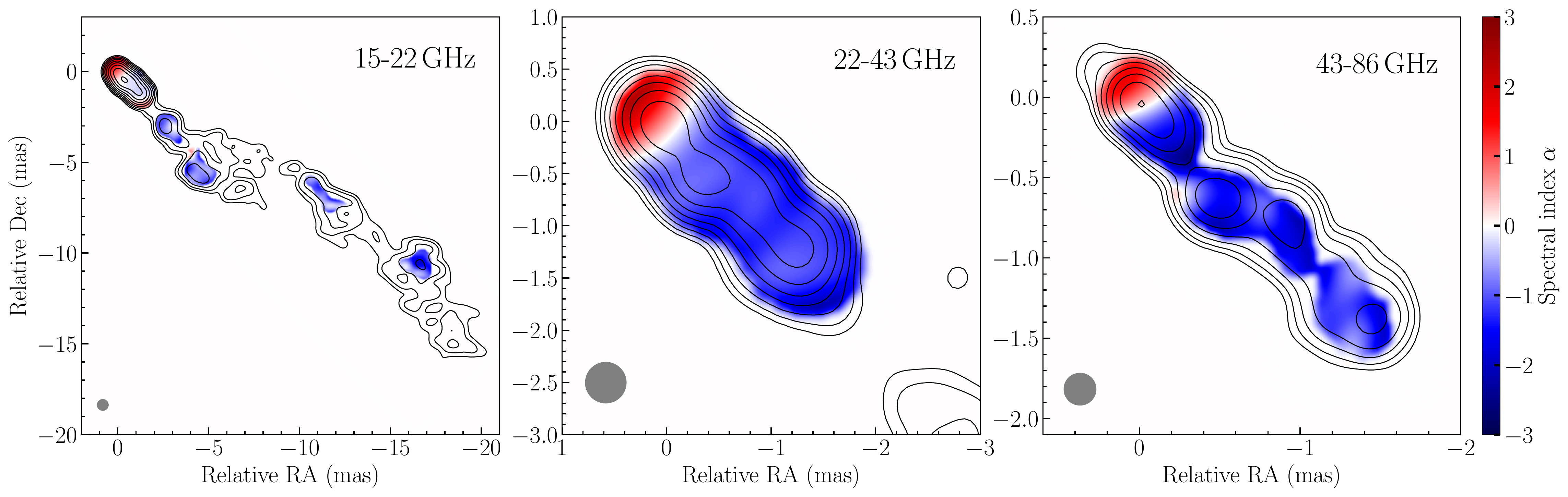}
 \caption{Spectral index maps between proximity frequency pairs labeled in the top right corner of each panel. Spectral index $\alpha$ is defined as $I_\nu \propto \nu^\alpha$. Spectral index values in each pair were calculated from all images in the top-sets of both frequencies after aligning the image using cross-correlation (Section\,\ref{sec:core_measure}) and averaging them. The contours in each panel represent the mean image of the lower frequency of the HSA as shown in Figure\,\ref{fig:HSA&VLBA_mean_image}.}
 \label{fig:index_map}
\end{figure*}

We measured the spectral index (defined $\alpha$ as $I_\nu\propto\nu^\alpha$) of the jet emission between two adjacent frequency bands. To derive the spectral index map for each pair of images, we first aligned the two images by adopting the offset determined by the two-dimensional cross-correlation (see Section\,\ref{sec:core_measure}). Then, the spectral index maps were derived for all possible pairs of top-set images at two frequencies. Then, we averaged over all spectral index maps for each frequency pair to create the mean spectral index map.

In Figure\,\ref{fig:index_map}, we show the mean spectral index map of the 3C\,273 jet for each pair of neighboring frequency bands. In all frequency pairs, the 3C\,273 jet consistently shows optically thick spectra upstream of the core, turns into flat spectra near the core, and then becomes optically thin downstream. These features are commonly observed in radio jet sources \citep[e.g.,][]{Hovatta14}, therefore supporting that the core offsets derived in Section\,\ref{sec:core_measure} are plausible.

\subsection{Core shift} \label{sec:core_shift}

\begin{table}[ttt]
 \begin{center}
 \caption{The positional offsets of the core between different frequencies}
 \label{tab:result_core_shift}
 \begin{tabular}{ccccccc}
    \hline \hline
    $\nu_1,\,\nu_2$ & N & pixel & $\mu_x$ & $\mu_y$ & $\sigma_x$& $\sigma_y$\\
    (GHz) &  & (mas) & (mas) & (mas) & (mas) & (mas)\\
    (1) & (2) & (3) & (4) & (5) & (6) & (7) \\
    \hline
    15.4, 23.7 & 1350 & 0.032 &  0.000 & 0.036 & 0.032 & 0.032 \\
    23.8, 43.2 & 1050 & 0.020 & -0.224 & 0.005 & 0.020 & 0.020 \\
    15.4, 43.2 & 1575 & 0.032 & -0.129 & 0.032 & 0.032 & 0.032 \\
    43.2, 86.3 & 840  & 0.010 & -0.077 &-0.010 & 0.010 & 0.010 \\
    \hline
 \end{tabular}
 \end{center}
 Notes: (1) Frequency pairs; (2) Total number of combinations;
 (3) Pixel size of re-gridded images used to measure offsets; (4) Mean offset in $x$-direction; (5) Mean offset in $y$-direction; (6) Standard deviation of the offsets in $x$-direction; (7) Standard deviation of the offset in $y$-direction. 
\end{table}

In Table\,\ref{tab:result_core_shift}, we show the measured core offsets for each combination of frequency pairs. The measured offsets are along $x$-direction, parallel to the large-scale jet axis.
The standard deviations of the measured offsets are the same or less than the pixel sizes of the re-gridded images used for these measurements. Therefore, the measured offsets do not depend on the different combinations of imaging parameter sets used to reconstruct the top-set images. 

We fit a power-law model of the core shift caused by the synchrotron self–absorption of the jet emission \citep[e.g.,][]{Lobanov98}. We assume that the core location at the frequency $\nu_i$ relative to the 86\,GHz core is given by
\begin{align}
    \label{eq_deltax} \Delta {x_i} &= A_x(\nu_i^{-k} -\nu_{86}^{-k}),\\
    \label{eq_deltay} \Delta {y_i} &= A_y(\nu_i^{-k} -\nu_{86}^{-k}),
\end{align}
for each direction, where $\nu_{86}=86.3$\,GHz. 
The power-law index $k$ depends on the optically thin spectral index of the synchrotron emission, magnetic field, and particle density distributions. Assuming a conical jet having a constant velocity and in the equipartition state between magnetic and particle energy densities, $k\approx1$ has been shown from the synchrotron self-absorption model proposed by \citet{Blandford79}, which was confirmed in many sources by previous core shift measurements \citep[see also][and references therein]{Pushkarev12}.
For 3C\,273, \citet{Lisakov17} has performed multi-epoch core shift measurements and reported values of $k$ close to $1$ for each epoch.
Therefore, given the small sets of frequency pairs available in the presented observations, we assume $k=1$ and only derive coefficients $(A_x, A_y)$ in our analysis. 

In this model, the positional offsets ($\mu_x$,\,$\mu_y$) of the radio cores between each frequency pair ($\nu_i$, $\nu_j$) are described as
\begin{align}
    \mu_x &= \Delta {x_i}-\Delta {x_j} = A_x(\nu_i^{-k} -\nu_j^{-k}),\\
    \mu_y &= \Delta {y_i}-\Delta {y_j} = A_y(\nu_i^{-k} -\nu_j^{-k}).
\end{align}
We derive the best-fit model parameters by comparing the model offsets above with measured ones through the least-squares fitting based on $\chi^2$ defined by
\begin{multline}
    \chi^2 = 
    \sum_{(i,j)} 
    \left\{\frac{A_x(\nu_i^{-k}-\nu_j^{-k})-\mu_x}{\sigma_x}\right\}^2\\             +\sum_{(i,j)}
    \left\{\frac{A_y(\nu_i^{-k}-\nu_j^{-k})-\mu_y}{\sigma_y} \right\}^2.
\end{multline}

\begin{figure}[ttt]
 \centering
 \includegraphics[width=1.0\columnwidth]{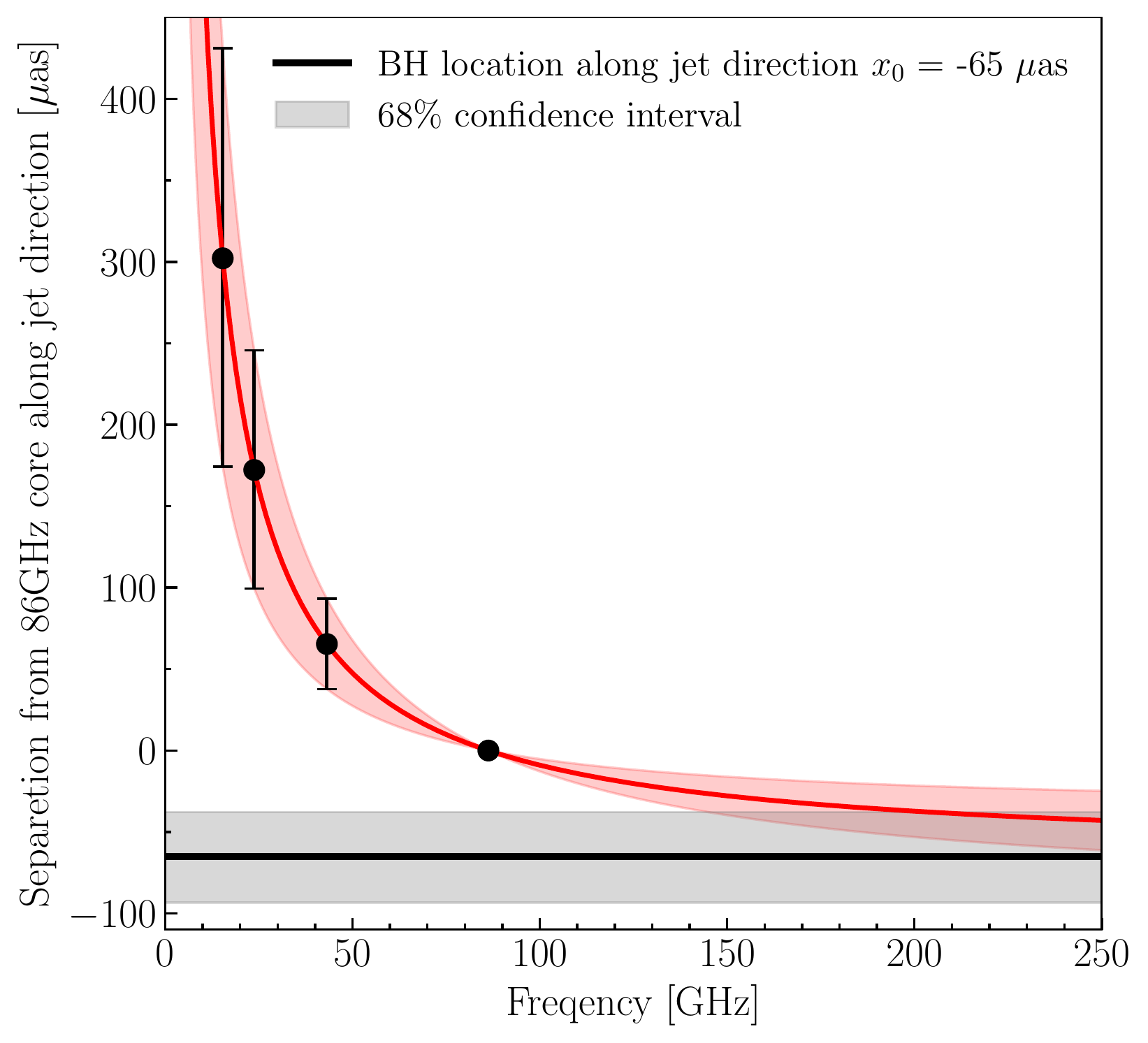}
 \caption{
 The core position along the large-scale jet axis as a function of the frequency based on the fitted power-law model. The core positions, denoted by the black dots, are relative to the 86\,GHz core. The red line shows the best-fit power-law model, and the black line shows the best-fit location of the central black hole. The error bars of the best-fit core positions and the shaded area of the best-fit model indicate the 68$\%$ confidence interval estimated from the $\chi^2$ surface of the fitted parameters (Section\,\ref{sec:core_shift}).}
 \label{fig:coreshift_jet}
\end{figure}

We found the best-fit parameters of $A_x=-5.635\pm2.424$ and $A_y=0.278\pm2.424$, which minimize $\chi^2$. The errors here indicate a $68.3\%$ confidence interval derived from the $\chi^2$ surface. With these values, Equations\,(\ref{eq_deltax}) and (\ref{eq_deltay}) can be used to derive the location of the upstream end of the jet, often considered the black hole position \citep[e.g.][]{Hada11}, by taking $\nu_{i}\rightarrow \infty$. From the best-fit parameters, we obtained the location of the jet apex at $x_0=65\,\rm{\mu as}$ and $y_0=-3\,\rm{\mu as}$ upstream\footnote{The location of the jet apex was derived for $k=1$, and may vary by at most a factor of two in the range of $k\sim0.8-1.2$ constrained by \citet{Lisakov17}. These uncertainties do not affect the main conclusions of this paper.} of the 86\,GHz core for 3C\,273.
Figure\,\ref{fig:coreshift_jet} shows the resultant core positions along the large-scale jet axis (i.e. $x$-direction).
The lower-frequency core is located at a farther distance from the central black hole, as anticipated for the core shift caused by synchrotron self-absorption \citep{Blandford79, Lobanov98}.

\subsection{Jet collimation profile of 3C 273} \label{sec:jet_collimation}

\begin{figure*}[ttt]
 \centering
 \includegraphics[width=0.84\textwidth]{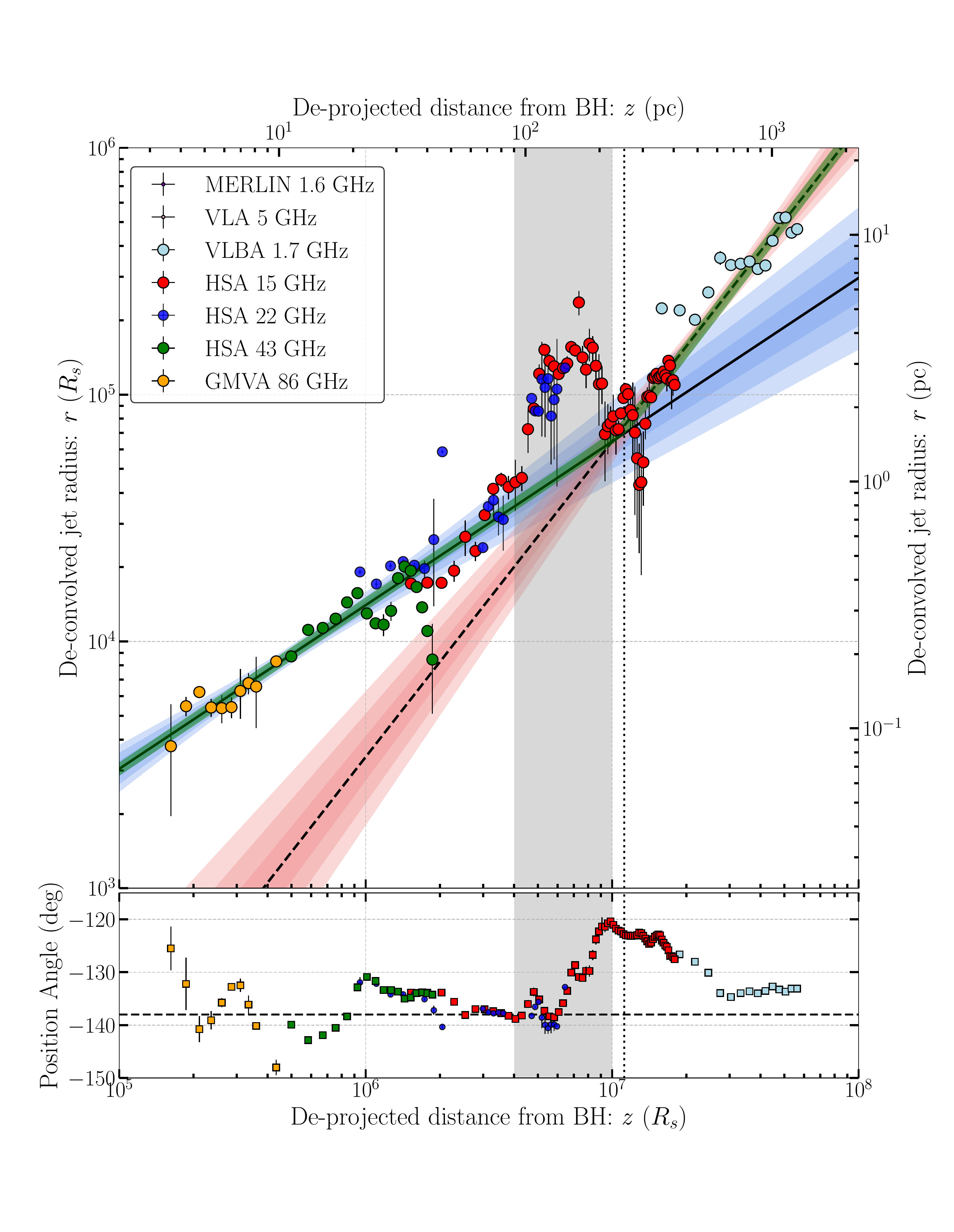}
 \vspace{-5em}
 \caption{
 The radius and PA of the 3C\,273 jet as a function of the de-projected distance from the central black hole. The de-projected distance from the central black hole is derived using the best-fit core shift model at a viewing angle of $i=9^{\circ}$ (see Section\,\ref{sec:intro}) for the black hole mass of $M_{BH}=2.6\times10^8 M_{\odot}$ \citep{GRAVITY18}. (\textit{The upper panel}): the jet radius profile. The colored circles and associated error bars indicate the mean values and standard deviations of the measurements, respectively. Different colors show measurements from different observations, as shown in the legend. The solid and dashed lines show single power-law fits ($r \propto z^a$) to the inner side, semi-parabolic ($a=0.66$), and to the outer side, conical/hyperbolic ($a=1.28$), respectively. The surrounding shaded areas represent the uncertainty of the single power-law fits ($1, 2, 3\times\sigma$). The green line represents the best-fit broken power-law function, giving the best-fit distance of the jet shape break of $z_b = 1.1\times10^7\,\Rs$ shown on the vertical dotted line. 
 (\textit{The lower panel}): the jet PA profile. The color convention is the same as in the upper panel. The horizontal dashed line shows ${\rm PA}= -138^{\circ}$, which is broadly consistent with the axis of the large-scale jet \citep[e.g.,][]{Davis85, Conway93}. The gray shade across both panels shows the region excluded from the fitting analysis because of a large bending seen in the jet (see the details in Section\,\ref{sec:jet_collimation}).}
 \label{fig:collimation}
\end{figure*}

Figure\,\ref{fig:collimation} shows the radius and position angle profile of the 3C\,273 jet as a function of the deprojected distance from the central black hole. Our VLBI measurements cover the de-projected distances ranging from $\sim 10^5\,\Rs$ to $\sim 10^{8}\,\Rs$. The radius profile at each frequency smoothly connects. Near $z \sim 10^7\,\Rs$, the jet radius profile shows a sharp increase by a factor of $\sim2$, where the PA of the jet sharply changes by $\sim20^{\circ}$ north.
The overall PA profile meanders around the mean PA of $-138^{\circ}$, coinciding with the large-scale jet \citep{Conway93}.

The jet radius profile at the outer region greater than $\sim 10^{7}\,\Rs$ is steeper than that of the inner region. Thus, we fit the power-law functions to the jet radius profile to quantify the jet shape in each region. The details of our model selection are described in  Appendix\,\ref{sec:appenndA}.

\begin{figure}
\centering
 \includegraphics[width=\columnwidth]{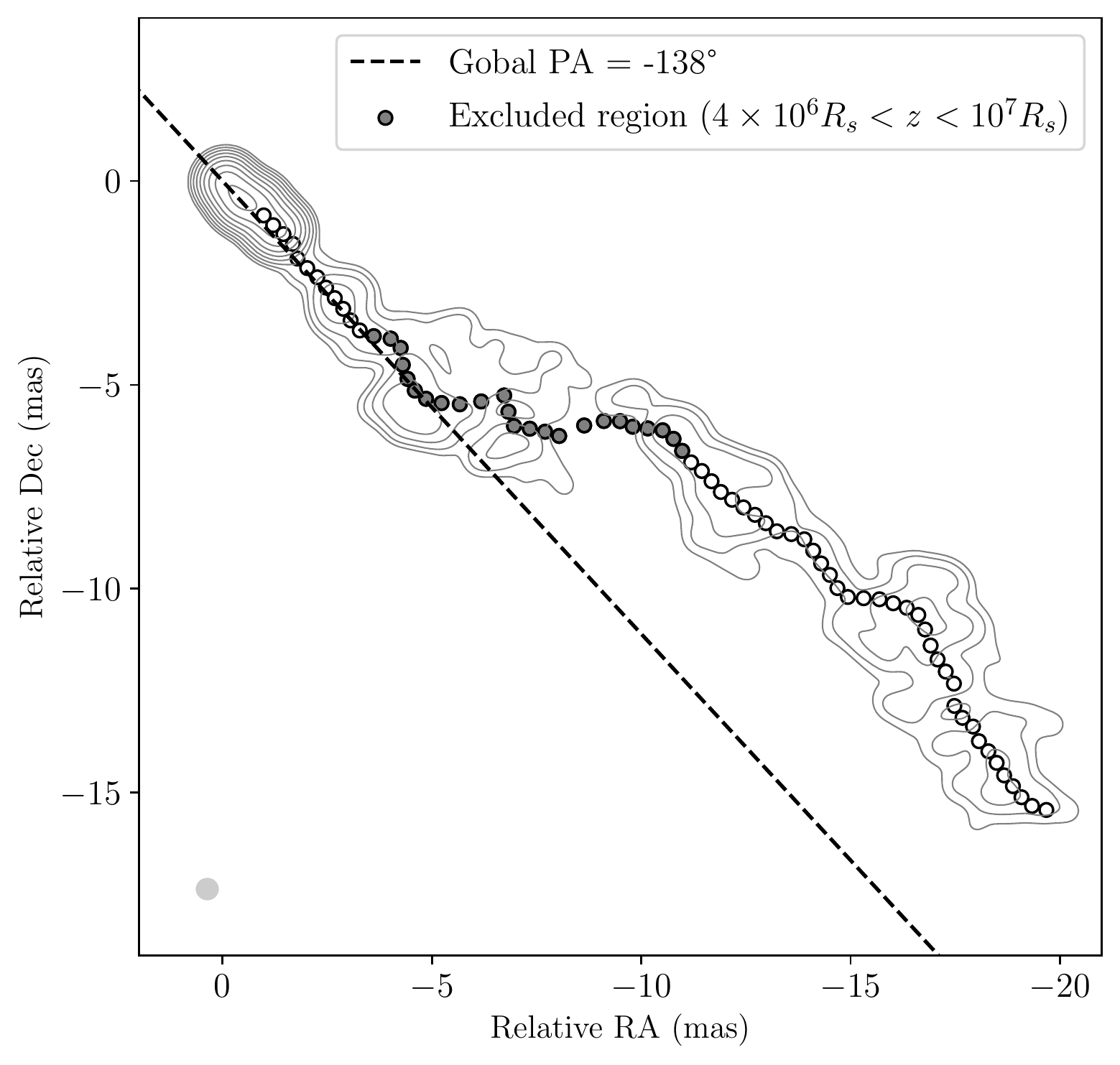}
 \caption{One of the top-set images from HSA 15\,GHz observation (gray contours) overlaid with the jet streamline of the 3C\,273 jet (filled and open circles). The excluded regions from our fitting are indicated by gray-filled circles, showing a stable complex and bending double-ridge-like morphology not seen in the rest of the global jet (see Section\,\ref{sec:colli_nature}). A black dashed line indicates the global position angle of the jet estimated by the large scale jet image \citep[e.g.,][]{Davis85}.}
 \label{fig:jet_ridge_appendix}
\end{figure}

Here, we excluded the region from $4\times10^6\,\Rs$ to $10^7\,\Rs$ (the gray shaded region in Figure \ref{fig:collimation}) from our fitting analysis because of its peculiar and complex morphologies not seen in the rest of the jet. In Figure\,\ref{fig:jet_ridge_appendix}, we show the streamline of the jet image of HSA 15\,GHz to illustrate the region that we excluded in our fitting analysis. This region shows a rapid swing in the streamline from the global position angle, and simultaneously the vertical profiles of the jet show highly asymmetric ridge-like structures not seen in the rest of the global jet. These peculiar morphologies make it difficult to define and measure the width of the jet in a consistent way with the rest of the global jet.

First, we model the jet radius $r(z)$ with a couple of single power-law functions ($r(z)\propto z^{a}$), separately fitting to each of the upstream ($z<4\times10^{6}\,\Rs$) and downstream ($>10^{7}\,\Rs$) regions.
The best-fit values and associated uncertainties of the parameters are estimated using the percentile bootstrap method, where in each trial the parameters are derived by unweighted least-squares method to the data sets randomly re-sampled from the original data allowing duplication. After 10,000 times bootstrap trials, we adopt the median values as the best-fit parameters, and one-third of the $99.7\,\%$ percentile confidence intervals as their $1\sigma$ uncertainties. The power-law index is found to be $a=1.28^{+0.08}_{-0.05}$, indicating a conical/hyperbolic shape in the downstream region, whereas $a=0.66^{+0.04}_{-0.05}$ is indicative of parabolic collimation found in the upstream region.

Furthermore, we also examine a joint fitting of both regions with a broken power-law function defined by
\begin{equation} 
    \label{eq:broken}
    r(z) =
    R_0\,2^{\frac{a_u-a_d}{2}}
    \left[ 1+ \left(\frac{z}{z_b} \right) \right]^{\frac{a_d-a_u}{s}}
\end{equation}
following \citet{Nakahara18}. $R_0$ is the scale factor, $z_b$ is the location of the break, and $a_{\rm u}$ and $a_{\rm d}$ are the power-law indices of the upstream and downstream regions, respectively.
Here $s$ is the parameter controlling the sharpness of the curvature at the transition point, which is fixed to be $s=10$. The best-fit parameters and their errors are derived using the bootstrapping method in the same manner. The power-law indices in the downstream and upstream regions are found to be $a_{\rm d}=1.31^{+0.10}_{-0.06}$ and $a_{\rm u}=0.67^{+0.02}_{-0.04}$, respectively, which are both consistent with the results of the above double power-law fitting. The jet shape transition is found to be $z_{\rm b} = 1.1^{+0.1}_{-0.2}\times 10^{7}\,\Rs$. These results show that the 3C\,273 jet has a transition in its shape from semi-parabolic to conical/hyperbolic at the deprojected distance of $\sim 10^{7}\,\Rs$ from the central black hole.

\section{Discussion} \label{sec:discuss}

\subsection{Nature of the Jet Collimation in 3C\,273} \label{sec:colli_nature}

\begin{figure}[ttt]
 \centering
 \includegraphics[width=1.0\columnwidth]{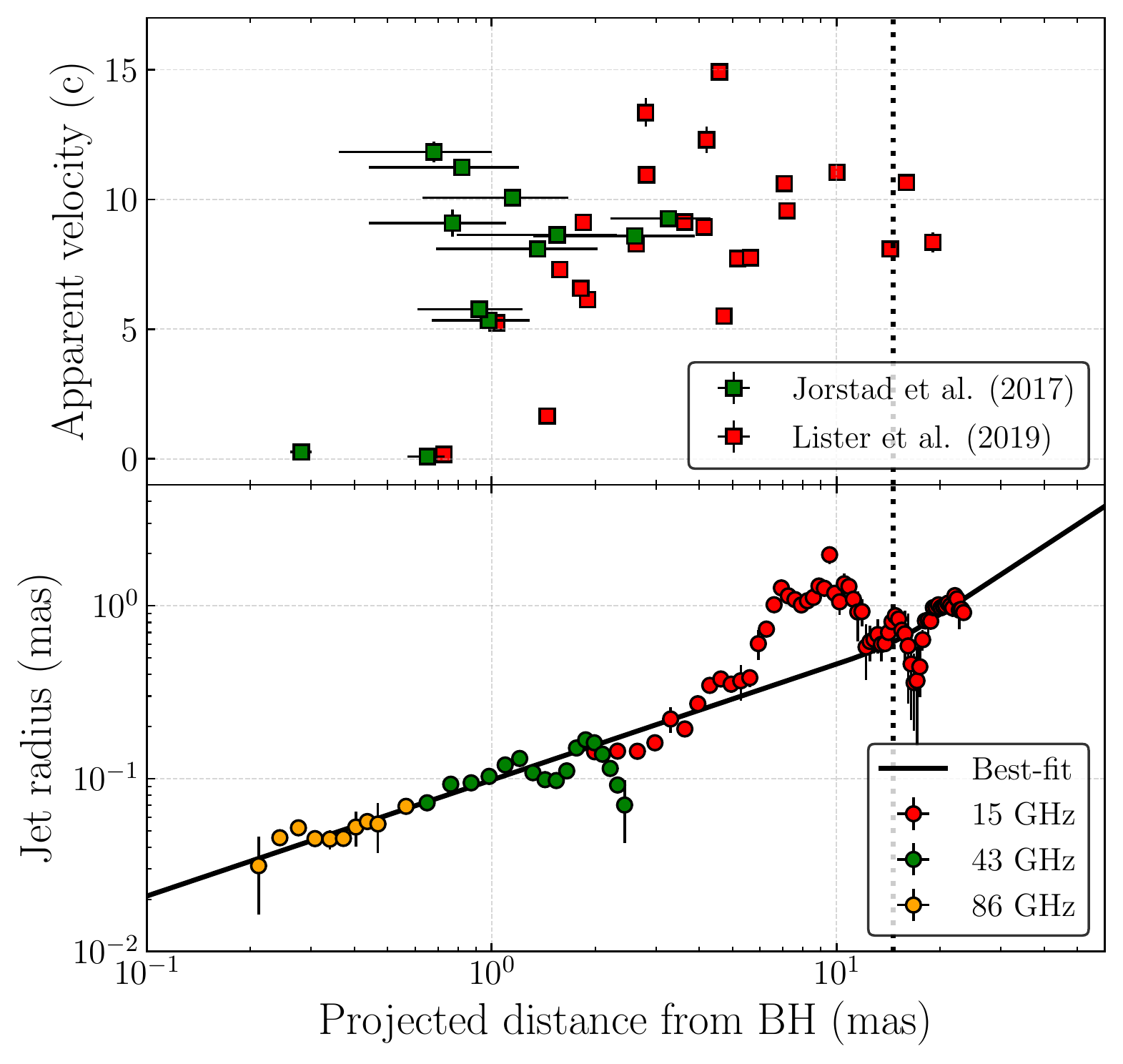}
 \caption{The apparent velocity (top), and collimation profile (bottom) as a function of the projected distance from the black hole. The black dotted line represents the best-fit location of the collimation break.}
 \label{fig:colli_velo}
\end{figure}

\begin{figure}[ttt]
 \centering
 \includegraphics[width=1.0\columnwidth]{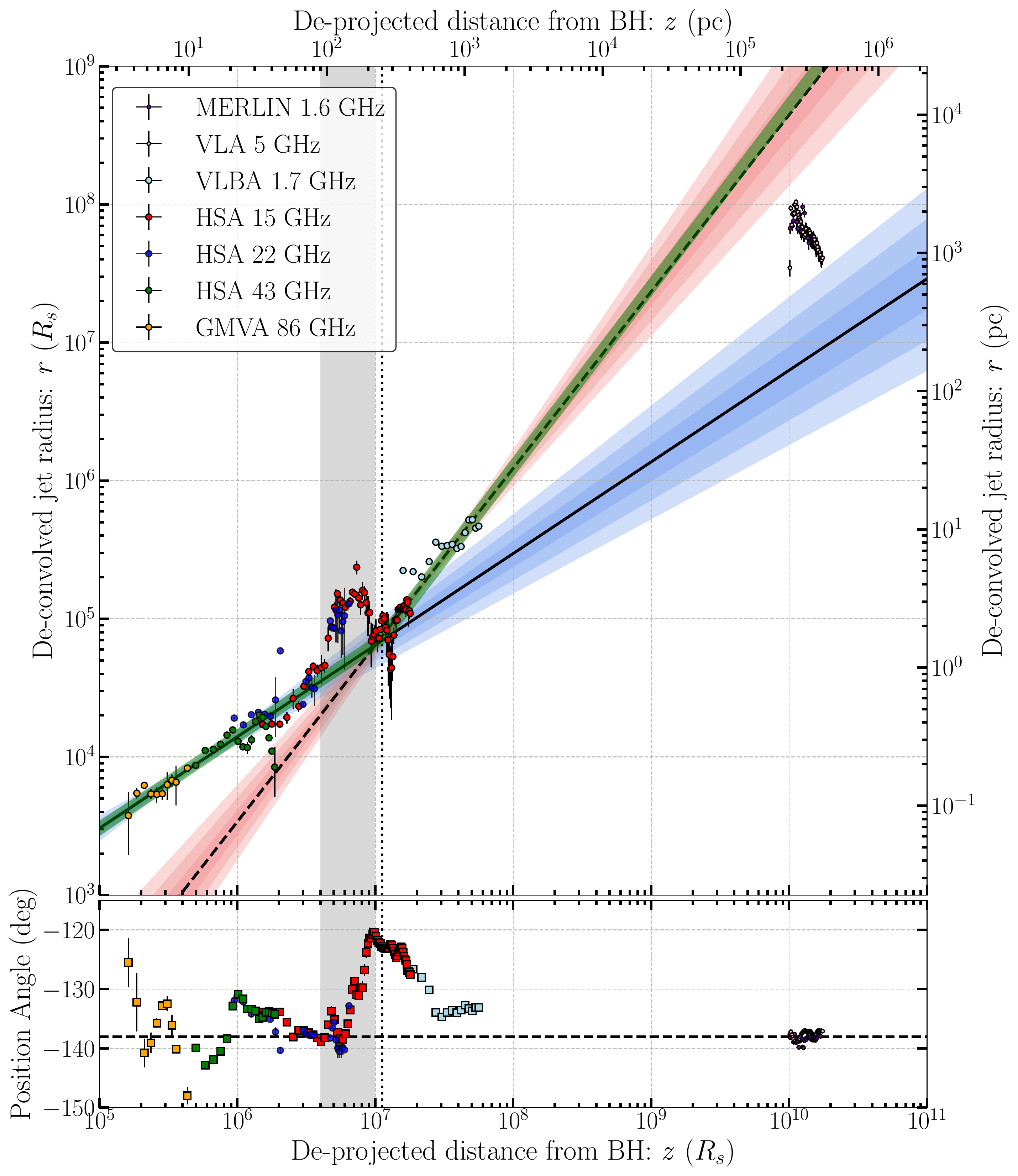}
 \caption{The jet radius and PA profiles of the 3C\,273 jet, extended from Figure\,\ref{fig:collimation} to include measurements of the large-scale jet beyond the deprojected distance of $\sim$200\,kpc traced with VLA 5\,GHz and MERLIN 1.6\,GHz.}
 \label{fig:collimation_large}
\end{figure}

Our new quasi-simultaneous observations clearly show the transition of the jet shape from semi-parabolic to conical/hyperbolic at $\sim 10^7\,R_s$ from the central black hole, providing the first compelling example of a collimation break in a quasar jet.
Interestingly, the transitional location roughly coincides with the region at $\sim 4\times10^6-10^7\,\Rs$ where the jet width ($2r$) rapidly expands by a factor of $\sim2$ along the significant bending, which was not used in the fitting because of its peculiarity (see Figure\,\ref{fig:jet_ridge_appendix}).
Recent space VLBI observations with Radio Astron at 1.6\,GHz revealed the complicated limb-brightened jet emission in this area \citep{Bruni21}.
\citet{Mertens15} shows non-ballistic motions of jet components along the northern and southern limbs over 14\,years, which is longer than the jet crossing time of this region considering the typical proper motion of the jet ($\sim1$\,mas/year, \citealt{Lister19}).  
After passing through the location of the collimation break, the northern and southern limbs smoothly merge into a single limb, and its position angle farther downstream is timely and spatially stable over 26\,years of 15\,GHz observations of the MOJAVE program \footnote{\url{http://www.physics.purdue.edu/astro/MOJAVE/sourcepages/1226+023.shtml}}. 
The stability of the broadened, double-ridge structure, apparently around the collimation break, suggests a stationary magnetohydrodynamic feature, possibly triggered by some changes in the circum-jet environment causing pressure mismatch with the jet \citep[e.g.,][]{Mizuno15}, as the HST-1 knot in M87 near the collimation break is indicated to be \citep{asada_nakamura12}.

The jet velocity field is another important physical quantity that is closely related to jet collimation.
This prediction is supported by various VLBI observations of M87 \citep{nakamura_asada13, Asada14, Park19b, Park21}, 1H\,0323+342 \citep{Hada18}, and statistical studies of large samples of AGN jets \citep{Homan15}.
In Figure\,\ref{fig:colli_velo}, we show the profile of the apparent jet velocities of 3C\,273 from long-term monitoring observations at 15\,GHz by the MOJAVE \citep{Lister19} and 43\,GHz by the BU-Blazar \citep{Jorstad17} programs, compared to the collimation profile.
Except for the stationary knot features at $\lesssim$1\,mas identified in the literature \citep{Savolainen06, Jorstad05, Jorstad17, Lisakov17, Bruni17}, Figure\,\ref{fig:colli_velo} indicates that the jet has already been accelerated to relativistic speed with an apparent velocity of at least $\sim5c$ before the collimation break. This trend in the 3C\,273 jet differs from that of M87 showing the gradual accelerations in the parabolic regions and velocity saturation at the collimation break \citep{Asada14, Park19a}.
Further observations of the jet kinematics in both the inner ($<1$\,mas) and outer regions ($>10$\,mas) would be useful for investigating the detailed velocity field in the collimation zone of 3C\,273, providing useful comparisons to radio galaxies.

For the inner-most jet at the inner sub-${\rm pc}$ scale, one can compare our results with recent sub-milliarcsecond-scale observations with near-infrared optical interferometry.
3C\,273 has been intensively studied with the Very Large Telescope Interferometer (VLTI), which recently resolved the broad line regions (BLRs) with a size of $46\,{\rm \mu as} \sim 0.1\,{\rm pc}$ as an ionized gas disk \citep{GRAVITY18}, and the dust torus emission with a size of $\sim150\,{\rm \mu as} \sim 0.4\,{\rm pc}$ \citep{GRAVITY20}. 
The innermost collimation profile of the 3C\,273 jets, expected by extrapolating to a farther inner region, is smaller than the measured geometry of the resolved BLRs and hot dust torus distributions. Such a region, where the jet is surrounded by the BLRs and dust torus, would be reachable with future higher frequency observations, such as with the EHT at 230 or 345\,GHz. Future higher-frequency VLBI observations will be critical for understanding the role of such an external medium common in quasars in jet collimation.

Downstream of the collimation break, the jet structure up to the $\sim$kpc scale shows a conical/hyperbolic expansion with a power-law index of $a=1.28$. These unconfined structures downstream of the jets have been found in other AGN jets \citep[][]{Pushkarev17, Kovalev20}.
To investigate the jet shape beyond the VLBI scales ($>\sim1$\,kpc), we show additional measurements based on CLEAN images from MERLIN and VLA observations at 1.6\,GHz and 5\,GHz in \citet{Akiyama18} and \citet{Perley17}, respectively, in Figure\,\ref{fig:collimation_large}.
These additional measurements are obtained from the kpc-scale radio lobe at the terminal end of the 3C\,273 jet.
Although previous observations detected the intermediate-scale ($\sim1-10\,{\rm arcsec}$) jet emission between the central core and the kpc-scale lobe \citep[e.g.,][]{Perley17}, the transverse structure is not well resolved, and the radius cannot be measured reliably. This causes a lack of measurements in $10^{8}-10^{10}\,\Rs$.
The inner semi-parabolic line, if extrapolated, significantly underestimates the jet radius at the kpc-scale. On the other hand, the jet radius at the kpc-scale is smaller than the extrapolated conical/hyperbolic line, and these kpc-scale jets can be more conically connected to 1.7\,GHz measurements.
Future observations of the intermediate scale would be useful for constraining the propagation of the jet into the kpc-scale radio lobe, and investigating detailed jet features interacting with the ambient ionized/molecular gas \citep{Husemann19} and hot gas cocoons \citep{Bromberg11}.

\subsection{Comparison with Other AGN sources} \label{sec:comparison}

\begin{figure*}[ttt]
 \centering
 \includegraphics[width=0.95\textwidth]{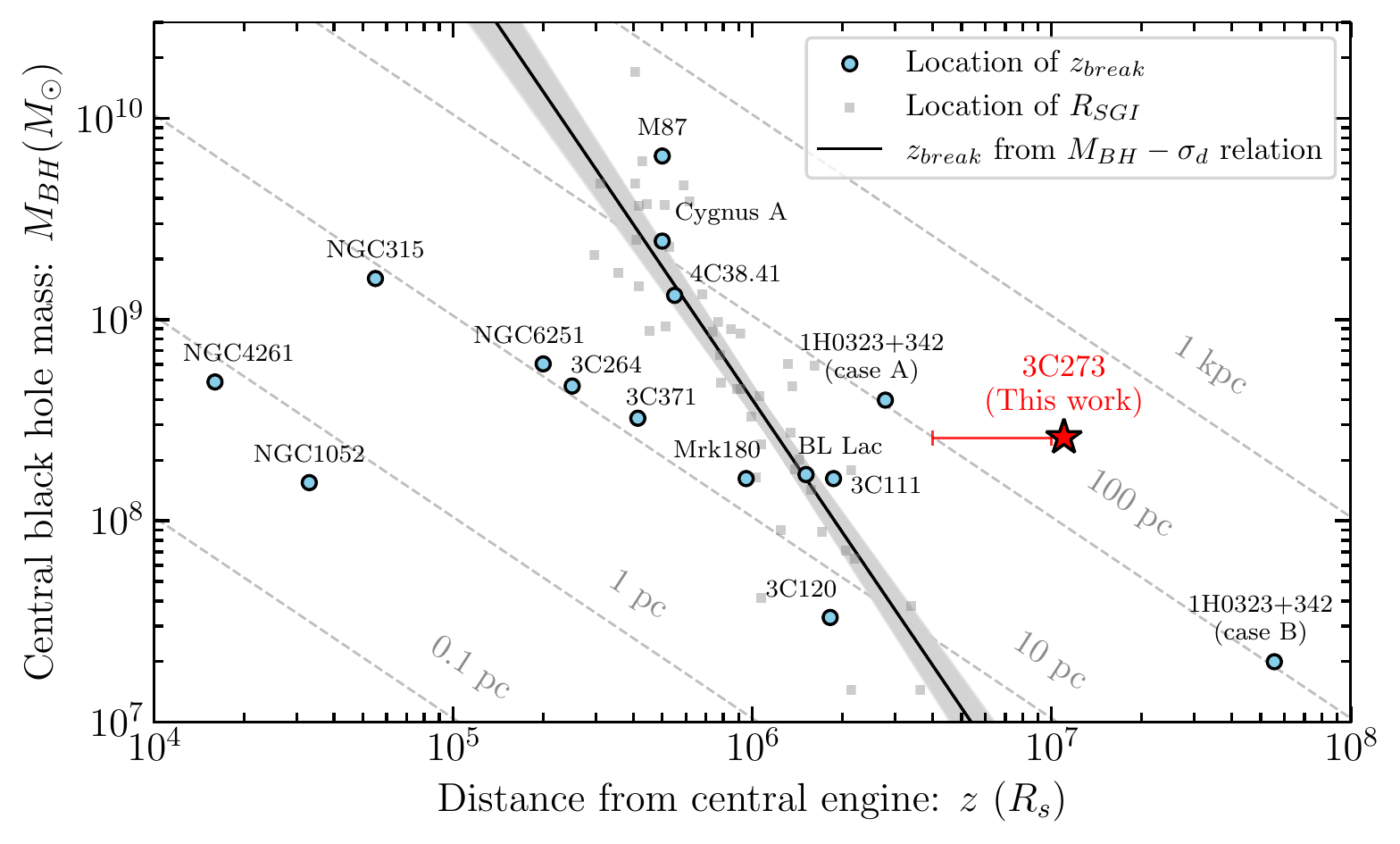}
 \caption{The relation between the black hole mass and the de-projected distance of the jet collimation break from the central engine. The red star shows our results from the 3C\,273 jet, and the cyan circles show results from other sources reported in the literature. The red horizontal bar indicates the region of a persistent feature seen near the break distance with a complex bending double-ridge-like morphology potentially associated with the transition in the jet shape (see Figure\,\ref{fig:jet_ridge_appendix} and Section\,\ref{sec:colli_nature}). To compare with the location of the sphere of the gravitational influence (SGI), we show the SGI locations of the classical bulges and elliptical galaxies in gray squares taken from \citet{KH13}. The black solid line and gray shaded area are the best-fit model and the 1-$\sigma$ error range of the relation between the black hole mass and SGI location, transformed from the $M_{\rm BH}-\sigma_{\rm d}$ relation in \citet{KH13}. The dotted lines show the actual distances in parsecs. The results of other sources are obtained from the following references; M87: \citealt{asada_nakamura12}, NGC\,6251: \citealt{Tseng16}, NGC\,4261: \citealt{Nakahara18}, Cygnus\,A: \citealt{Nakahara19_cygA}, NGC\,1052: \citealt{Nakahara20}, NGC\,315: \citealt{Park21}, 4C38.41: \citealt{Algaba19}, 3C\,264: \citealt{Boccardi19}, 1H\,0323+342: \citealt{Hada18}, and other sources: \citealt{Kovalev20}. Note that we show two cases of different black hole masses for 1H\,0323+342 (see Section\,\ref{sec:comparison}). 
}
 \label{fig:comparisons}
\end{figure*}

The jet structure has been investigated for many objects in the last decades, and the transitions of jet collimation profiles have been discovered in several jets of radio galaxies and BL Lac objects. The presented results on 3C\,273 have added a case as a quasar jet. To compare our 3C\,273 results with other AGN jets, in Figure\,\ref{fig:comparisons}, we show the relation between the black hole mass and the deprojected distance of the jet collimation break for both 3C\,273 and other AGN sources.
The break locations are widely distributed from $\sim10^4\,\Rs$ to $\sim10^8\,\Rs$, where 3C\,273 is located on the further side. 

A popular interpretation for the origin of the jet collimation break is that the transition of the external pressure profile triggers the end of the jet collimation, where the external pressure does not provide enough support to confine the jet \citep[e.g.,][]{asada_nakamura12, Tseng16}.
For the pioneering case of M87, \citet{asada_nakamura12} discussed that the transition of the jet shape may result from the transitional change in the external pressure profile of the circum-jet medium near the Bondi radius and the end of the sphere of gravitational influence (SGI).
This interpretation is motivated by X-ray observations, which indicate that the different pressure profiles of the ambient gas are realized inside and outside the Bondi/SGI radius in M87 \citep{Russell15, Russell18}.

We further investigated the hot gas properties for other sources to check whether similar conditions to M87 could be established. 
However, the number of sources is limited whose proximity to the central core can be spatially resolved by X-ray observations, making it difficult to give accurate estimations of the Bondi radii for those sources. Therefore, we consider the SGI radius instead of the Bondi radius for the gravitational spheres of the central black holes.
Note that we checked the temperature of the bright cores of several sources hosted by elliptical galaxies in our collected samples and confirmed that the Bondi radii are approximately of the same order of magnitude as the SGI radii \footnote{Considering the empirical M-$\sigma_{\rm d}$ relation for an AGN with a typical black hole mass ($10^{8-9}M_{\odot}$), we expect the temperature of the hot gas in the central core to be  $\sim 0.1-1$\,keV when the Bondi radius is close to the SGI radius.} \citep[e.g.,][]{Sun09, Fujita16}.

Here, the SGI radius is estimated as ${R_{\rm SGI}}=GM_{\rm BH}/\sigma_d^2=(1/2)(c/\sigma_d)^2\,\Rs$, where $\sigma_d$ is the stellar velocity dispersion.
To investigate the relationship between the location of the SGI boundary and the jet collimation break among our samples with various BH masses, we introduced an emprical $M_{\rm BH}-\sigma_d$ relation derived in \citet{KH13}. This relation is particularly strong for the classical bulge and elliptical galaxies, and our sample is supposed to follow this relation (e.g., see Table\,2 in \citealt{KH13}, and references therein). 
Indeed, 3C\,273 follows the relation from the stellar velocity dispersion of $\sigma_{\rm d} = 210$\,km/s \citep{Husemann19} and SMBH mass of $M_{\rm BH} = 2.6\times10^8\,M_{\odot}$, which gives us an estimation of the SGI radius as $R_{\rm SGI}\sim10^{6}\,\Rs$ (see the solid line in Figure\,\ref{fig:comparisons}).
 
The most important finding here is that the measured locations of the collimation breaks, including 3C\,273, are more widely distributed than the intrinsic scatter of the SGI locations seen in the samples of \citet{KH13}.
In particular, NGC\,4261 \citep{Nakahara18}, NGC\,1052 \citep{Nakahara20} and NGC\,315 \citep{Boccardi21, Park21} imply a transition of the jet collimation in a substantially inner area than the SGI boundary.
For NGC\,315, \citet{Park21} discussed that the non-relativistic outflow, such as wind from the accretion disk, may play a critical role in jet collimation --- for instance, the inner break of the jet collimation seen in NGC\,315 may occur if the wind does not reach the Bondi radius or the end of the SGI.

The other two sources have a common property at the jet base. They are surrounded by a dense obscuring disk/torus causing free-free absorption (FFA).
For NGC\,1052, the physical size of the FFA torus is $\sim 0.5\,{\rm pc}$ $(\sim 3.4\times10^4\,\Rs)$ in radius (\citealt{Kameno00}), which is close to the de-projected break location of $\sim0.15\,{\rm pc}$ reported from \citet{Nakahara20}\footnote{The break location is consistent with an independent measurement by \citet{Kovalev20} within a factor of $\sim 3$.}
For NGC\,4261, \citet{Haga15} estimated the transition radius in the accretion disk of $\sim2\times10^3\,\Rs$ from FFA measurements, which is close to the jet shape break of $\sim 10^4\,\Rs$ for NGC\,4261 \citep{Nakahara18}.

The break location of 1H\,0323+342 is not well constrained because of the significantly uncertain black hole mass. It shows a distant collimation break for the lower black hole mass (case\,B in Figure\,\ref{fig:comparisons}), while the jet collimation break occurs near the SGI for the larger black hole mass (case\,A). \citet{Hada18} discussed that radiation driven outflow/wind associated with the narrow-line region (NLR) may confine the jet of 1H\,0323+342 (case\,B), while 1H\,0323+342 and M87 may share the common jet collimation mechanism (case\,A), which is supported by 1H\,0323+342 having a stationary recollimation shock at the transitional region like the HST-1 knot in M87 \citep[see][]{Doi18}.

Our results and Figure \ref{fig:comparisons} indicate that the transition distance of the jet shape is not necessarily determined by the SGI boundary, but rather by its diverse environment, such as the presence of the disk, torus, or disk wind and their spatial extent.
The hot gas cocoon surrounding the jet may also be an influence \citep[e.g.,][]{Bromberg11}.

Recently, \citet{Boccardi21} discussed the jet collimation properties of various radio galaxies with respect to the activity of the central accretion disk by categorizing samples into low- and high-excitation radio galaxies (LERG and HERG).
They found that HERGs had larger radii and longer shape transition distances for the collimation properties.
For 3C\,273, it is unlikely that a component near the central engine, such as the torus, is the critical factor causing the break because the collimation break in the 3C\,273 jet is outside the range of $\sim100$\,pc. Then 3C\,273, if viewed from a typical viewing angle of radio galaxies, would be categorized as a HERG given the high ratio of the X-ray luminosity $L_{\rm X}$ ($2-10\,{\rm keV}$) to the Eddington luminosity $L_{\rm Edd}$ implied by $L_{\rm X}>1.4\times10^{46}\,{\rm erg/s}$ \citep{Cappi98} and $L_{\rm Edd}=3.4\times10^{46}\,{\rm erg/s}$ given for $M_{\rm BH}=2.6\times10^{8}\,M_{\odot}$ \citep{GRAVITY18}.
Therefore, our results of 3C\,273, which show relatively large jet widths ($2r>\sim10^{4}\,\Rs$ in $z>\sim10^{5}\,\Rs$) and distant location of the jet collimation break ($z_{\rm b}>10^{6}\,\Rs$), are consistent with the statistical trend reported in \citet{Boccardi21}. 

Finally, we briefly note a scenario in which the jet collimation break is governed by the internal jet physics proposed by \citet{Kovalev20}. They proposed that under a single ambient pressure profile, such as $p\propto z^{-2}$, the collimation break occurs in the region where the magnetic and particle energies are equivalent. In their model, the break location mainly depends on the initial magnetization parameters ($\sigma_{\rm M}$) at the jet base, as well as the black hole mass and spin \citep[see Equation\,7 in][]{Nokhrina20}.The large $\sigma_{\rm M}$ for 3C\,273 estimated in \citet{Nokhrina15} may be qualitatively consistent with the distant jet collimation break found in this study. However, it is still hard to tightly constrain the value of $\sigma_{\rm M}$, and it is still an open problem. Therefore, it remains unclear whether this scenario can explain the diverse distributions of the break locations ranging from $10^4\,\Rs$ to $10^8\,\Rs$ shown in Figure\,\ref{fig:comparisons}. If the value of $\sigma_{\rm M}$ can be estimated accurately for more sources, the physical connection between $\sigma_{\rm M}$ the break location could be discussed in more detail.

\section{Summary} \label{sec:summary}
In this paper, we have investigated the global jet structure of the archetypical quasar 3C\,273 with VLBA, HSA, and GMVA observations. In particular, we have reported on new GMVA observations at 86\,GHz conducted in the first session involving ALMA which significantly enhances the sensitivity and angular resolution of the images, providing the detailed morphology of the innermost jet. With the robust imaging analysis with the state-of-the-art RML imaging techniques, we obtained the jet collimation profile over a wide range of $10^5-10^{8}$\,$R_s$ in the deprojected distance. We summarize our results as follows.

\begin{itemize}
    \item The quasar 3C\,273 jet is found to have the structural transition from semi-parabolic to conical/hyperbolic shape at $\sim 10^{7}\,\Rs$, providing the first clear example of a quasar jet. Our results suggest the existence of the jet collimation break for sources exhibiting widely different accretion rates.
    \item The collimation break is located in the area at $\sim 4\times10^6-10^7\,\Rs$ where the jet shows a bending streamline for $\sim 10\,{\rm mas}$ persistent more than the typical jet crossing time. The area is also known as the broadened vertical structure with brightening two limbs resolved by recent Space VLBI observations. This peculiar and persistent feature, co-located with the jet break, may be explained by a stable magneto-hydrodynamic feature, as HST-1 may be for M87.
    \item The extrapolation of the collimation profile is consistent with the spatial distribution of BLRs and dusty winds resolved with recent GRAVITY observations. Our observations suggest that future higher-frequency observations, for instance, with the Event Horizon Telescope at 230 and 345\,GHz can probe further inner regions where the jet might interact with such hot dusty ionized gas.
    \item Comparison to the collimation profile of the 3C\,273 jet with the velocity field obtained by long-term monitoring observations at 15 and 43\,GHz shows that the 3C\,273 jet is already accelerated to its apparent maximum velocity at the upstream of the collimation break, which is apparently different from the M87 jet where the ends of the collimation and acceleration zones are co-located.
    \item The collimation break of 3C\,273 is located nearly several-to-ten times further than the estimated SGI location. Furthermore, we compared all available locations of the jet break from the literature with the SGI locations derived from $M_{\rm BH}-\sigma_{d}$ relation, clearly showing that the transition of the jet collimation is governed not only by the SGI of the central black hole.
    \item Relatively further location of the jet collimation break discovered for 3C\,273 seems broadly consistent with the statistical trend of radio galaxies that higher excitation AGN sources tend to end their jet collimation at more distant locations from the central black hole.
\end{itemize}
    
Our results demonstrate that ultra-high-resolution observations with new millimeter facilities can provide many new insights for high-powered AGN which were not previously suitable to study jet collimation.
Further study using full polarimetric data and Faraday rotation synthesis can reveal the distribution of the three-dimensional magnetic field and the circumnuclear medium in the active collimation region \citep[e.g.,][]{Park19a}.
These explorations with polarimetric analysis using presented multifrequency data including GMVA+ALMA will be presented in a forthcoming paper.

\newpage
\acknowledgments
We thank the anonymous referee for many helpful and constructive suggestions to improve this paper during the review stage.
This work was financially supported by grants from the National Science Foundation (NSF; AST-1440254; AST-1614868; AST-2034306) and the Japan Society for Promotion of Science (JSPS; JP21H01137; JP18H03721.).
This work has been partially supported by the Generalitat Valenciana GenT Project CIDEGENT/2018/021 and by the MICINN Research Project PID2019-108995GB-C22.
K.A. is also financially supported by the following NSF grants (AST-1935980, AST-2107681, AST-2132700, OMA-2029670).
S.I. is supported by NASA Hubble Fellowship grant HST-HF2-51482.001-A awarded by the Space Telescope Science Institute, which is operated by the Association of Universities for Research in Astronomy, Inc., for NASA, under contract NAS5-26555.
R.-S.L. is supported by the Max Planck Partner Group of the MPG and the CAS and acknowledges support by the Key Program of  the National Natural Science Foundation of China (grant No. 11933007), the Key Research Program of Frontier Sciences, CAS (grant No. ZDBS-LY-SLH011), and the Shanghai Pilot Program for Basic Research – Chinese Academy of Science, Shanghai Branch (JCYJ-SHFY-2022-013).
The Black Hole Initiative at Harvard University is financially supported by a grant from the John Templeton Foundation.
This paper makes use of the following ALMA data: ADS/JAO.ALMA2016.1.01216.V. ALMA is a partnership of ESO (representing its member states), NSF (USA), and NINS (Japan), together with NRC (Canada), MOST and ASIAA (Taiwan), and KASI (Republic of Korea), in cooperation with the Republic of Chile. The Joint ALMA Observatory is operated by ESO, AUI/NRAO and NAOJ.
This research has made use of data obtained with the Global Millimeter
VLBI Array (GMVA), which consists of telescopes operated by the (Max-Planck-Institut \"{f}ur Radioastronomie; MPIfR), IRAM, Onsala, Metsahovi, Yebes, the Korean VLBI Network, the Green Bank Observatory, and the Very Long Baseline Array (VLBA).
The VLBA is an instrument of the National Radio Astronomy Observatory.
The National Radio Astronomy Observatory is a facility of the National Science Foundation operated under cooperative agreement by Associated Universities, Inc. 
This work is partly based on observations with the 100-m telescope of the MPIfR at Effelsberg.
This work made use of the Swinburne University of Technology software correlator \citep{Dellar2011}, developed as part of the Australian Major National Research Facilities Programme and operated under licence.
This research has made use of data from the MOJAVE database that is maintained by the MOJAVE team \citep{Lister18}. 
This study makes use of 43\,GHz VLBA data from the VLBA-BU Blazar Monitoring Program (VLBA-BU-BLAZAR; \url{http://www.bu.edu/blazars/VLBAproject.html}), funded by NASA through the Fermi Guest Investigator Program.

\vspace{5mm}
\facilities{ALMA, VLBA, HSA, GMVA}

\software{
    SMILI \citep{Akiyama17a, Akiyama17b}, AIPS \citep{Greisen03}, Difmap \citep{Shepherd97}, DiFX \citep{Deller07}, astropy \citep{Astropy13, Astropy18}, numpy \citep{vanderWalt11}, scipy \citep{virtanen20}, matplotlib \citep{Hunter07}, pandas \citep{Pandas2010}.
}

\newpage
\appendix
\section{Model Selection} \label{sec:appenndA}

\begin{figure}[ttt]
 \centering
 \includegraphics[width=0.9\textwidth]{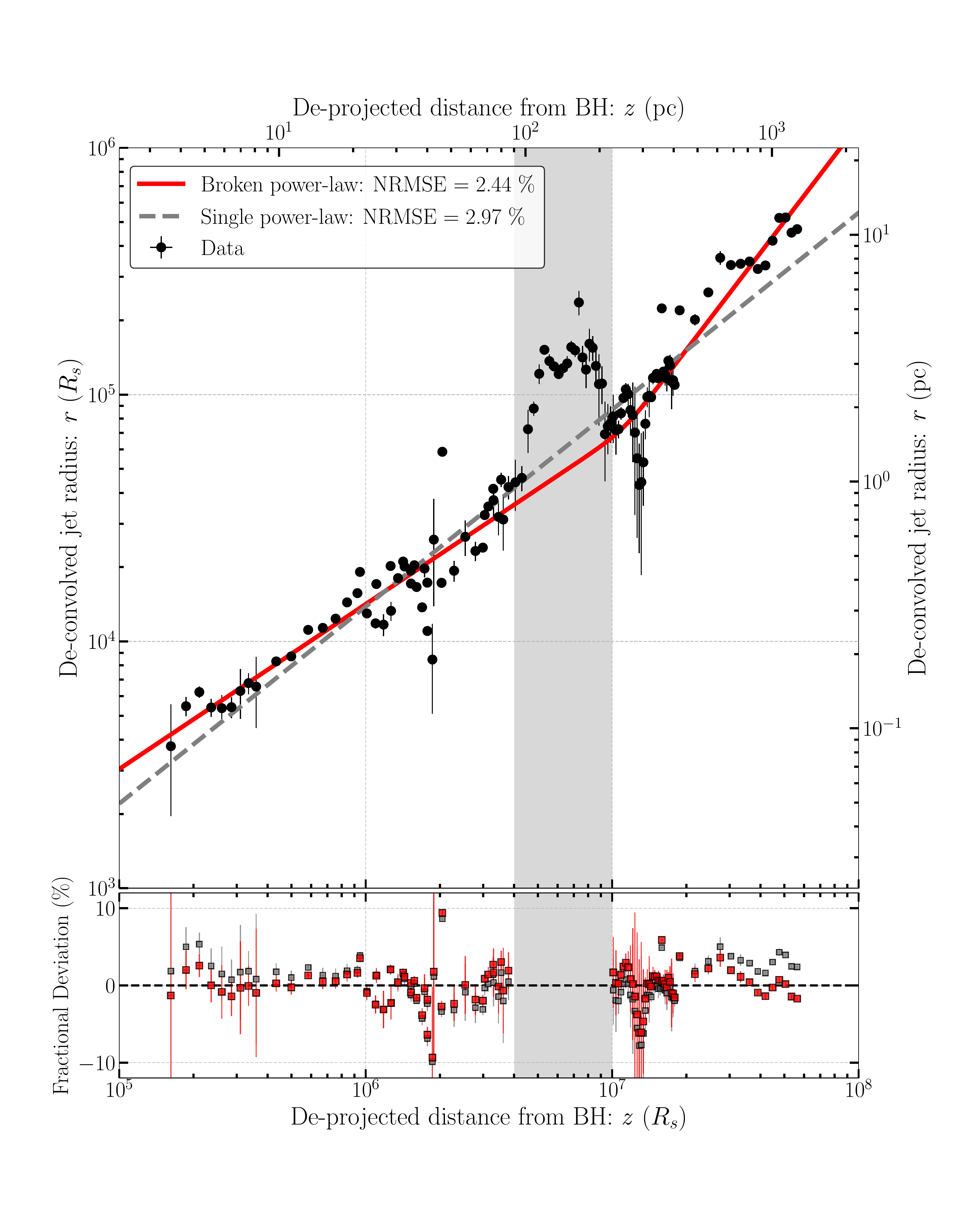}
 \caption{Upper panel: Radial profile of jet in 3C\,273 with different models of power-law functions. Measurements data are same as Figure\,\ref{fig:collimation}. Gray dashed line shows the best-fit single power-law function ($r \propto z^a$) with $a=0.80$ obtained from all the data except for the gray shaded region. The red solid line shows the best-fit broken power-law function as presented in Figure\,\ref{fig:collimation}. Lower panel: Fractional deviations of the jet radii from the obtained models. The red and gray-colored squares correspond to the deviations from the lines in the upper panel.}
 \label{fig:collimation_delta}
\end{figure}

We describe the model selection for our fitting analysis to the jet radius profile presented in Section\,\ref{sec:jet_collimation}. To choose a better model to explain the overall jet structure, we performed a fitting analysis using single and broken power-law functions for entire measurements except for the rapid increase of jet radii around $\sim10^{7}\,\Rs$ (see Section\,\ref{sec:jet_collimation}). Then we checked their fractional deviations $\delta$ defined as 
\begin{equation}
    \delta = \frac{r - r_{\rm model}}{r_{\rm model}}.
\end{equation}

For the case of a single power-law fit in Figure\,\ref{fig:collimation_delta}, the deviations from the model are seen in both inner and outer edges, while the relatively small for a broken power-law fit.

For more quantitative comparison, we computed the normalized mean square error (NRMSE) values \citep[e.g.,][]{Chael16, Akiyama17a, Akiyama17b} for each model, defined as 
\begin{equation}
    {\rm NRMSE} = \sqrt{\frac{\Sigma(r-r_{\rm model})^2}{\Sigma r^{2}_{\rm model}}}.
\end{equation}
As shown in Figure\,\ref{fig:collimation_delta}, the resultant NRMSE value for the broken power-law model is lower than that of the single power-law model, also supporting that broken power-law function is a better model describing the whole range of measured jet radii of 3C\,273.

\bibliography{bibtex}{}
\bibliographystyle{yahapj}

\end{document}